\documentclass[10pt,journal]{IEEEtran}
\usepackage[nocompress]{cite}
\usepackage{epsfig}
\usepackage{epstopdf}
\usepackage{graphicx}
\usepackage{float}
\usepackage{verbatim}
\usepackage{subcaption}
\usepackage{wrapfig}
\usepackage[procnumbered,ruled,vlined]{algorithm2e}
\usepackage[noend]{algorithmic}
\usepackage{amsmath}
\usepackage{amssymb}
\usepackage{multirow}
\usepackage[table]{xcolor}

\begin{document}

\title{Generation of Synthetic Spatially Embedded \\Power Grid Networks}

\author{Saleh~Soltan
        and~Gil~Zussman% <-this % stops a space
\IEEEcompsocitemizethanks{\IEEEcompsocthanksitem S. Soltan and G. Zussman are with the Department
of Electrical Engineering, Columbia University, New York,
NY, 10027.\protect\\
E-mail: \{saleh,gil\}@ee.columbia.edu}}

\maketitle
%\IEEEtitleabstractindextext{%
\begin{abstract}

The development of algorithms for enhancing the resilience and efficiency of the power grid requires performance evaluation with real topologies of power transmission networks. However, due to security reasons, \emph{such topologies and particularly the locations of the substations and the lines are usually not publicly available}. Therefore, we study the structural properties of the North American grids and present an algorithm for generating synthetic spatially embedded networks with similar properties to a given grid. The algorithm uses the Gaussian Mixture Model (GMM) for density estimation of the node positions and generates a set of nodes with similar spatial distribution to the nodes in a given network. Then, it uses two procedures, which are inspired by the historical evolution of the grids, to connect the nodes. The algorithm has several tunable parameters that allow generating grids similar to any given grid. Particularly, we apply it to the Western Interconnection (WI) and to grids that operate under the SERC Reliability Corporation (SERC) and the Florida Reliability Coordinating Council (FRCC), and show that it generates grids with similar structural and spatial properties to these grids. To the best of our knowledge, this is the first attempt to \emph{consider the spatial distribution of the nodes and lines} and its importance in generating synthetic power grids.
\end{abstract}

%}

\setlength{\textfloatsep}{4 pt}

%\IEEEdisplaynontitleabstractindextext

% \fi
\IEEEpeerreviewmaketitle

%\IEEEraisesectionheading{\section{Introduction}\label{sec:introduction}}
\section{Introduction}\label{sec:introduction}
The design of algorithms and methods for enhancing the power grid (namely, making it smarter) drew tremendous attention over the past decade~\cite{amin2008electric,fang2012smart}. These efforts focused on challenges stemming from renewable generation interconnection \cite{bienstock2014chance}, Phasor Measurement Units (PMUs) placement \cite{SYZ2015,zhao2012pmu}, transmission expansion planning \cite{latorre2003classification}, and vulnerability analysis \cite{Soltan2014Cascade,SmartGridComm11rep,asztalos2014cascading,chertkov2011predicting}.
The development of algorithms for coping with these challenges \emph{requires performance evaluation with real grid topologies}. However, in order to avoid exposing vulnerabilities, \emph{the topologies of the power transmission networks and particularly the locations of the substations and the lines are usually not publicly available} or are hard to obtain.

There are only very few and limited test cases and real-world power grid datasets that are publicly and freely available. These include the IEEE test cases~\cite{IEEEtestcase}, the National Grid UK~\cite{NGUK}, the Polish grid~\cite{Polishgrid}, and an approximate model of the European interconnected system~\cite{zhou2005approximate}. To the best of our knowledge, among these, National Grid UK is the only publicly available dataset with geographical locations.
Even if the data was available, it  would be unwise to publish vulnerability results which are based on real topologies, due to the enormous cost of grid enhancements. On the other hand, it was recently shown that simple random graph models cannot be used to generate grids with appropriate structural and spatial characteristics~\cite{cotilla2012comparing} (for more details, see Section~\ref{sec:related}).
Therefore, in this paper \emph{we design an algorithm for generating synthetic networks with similar structural and spatial properties to real power grids}. Such synthetic networks can be used for evaluation of various methods and techniques.

\begin{figure}[t]
\centering
\vspace*{-0.2cm}
\includegraphics[scale=0.25]{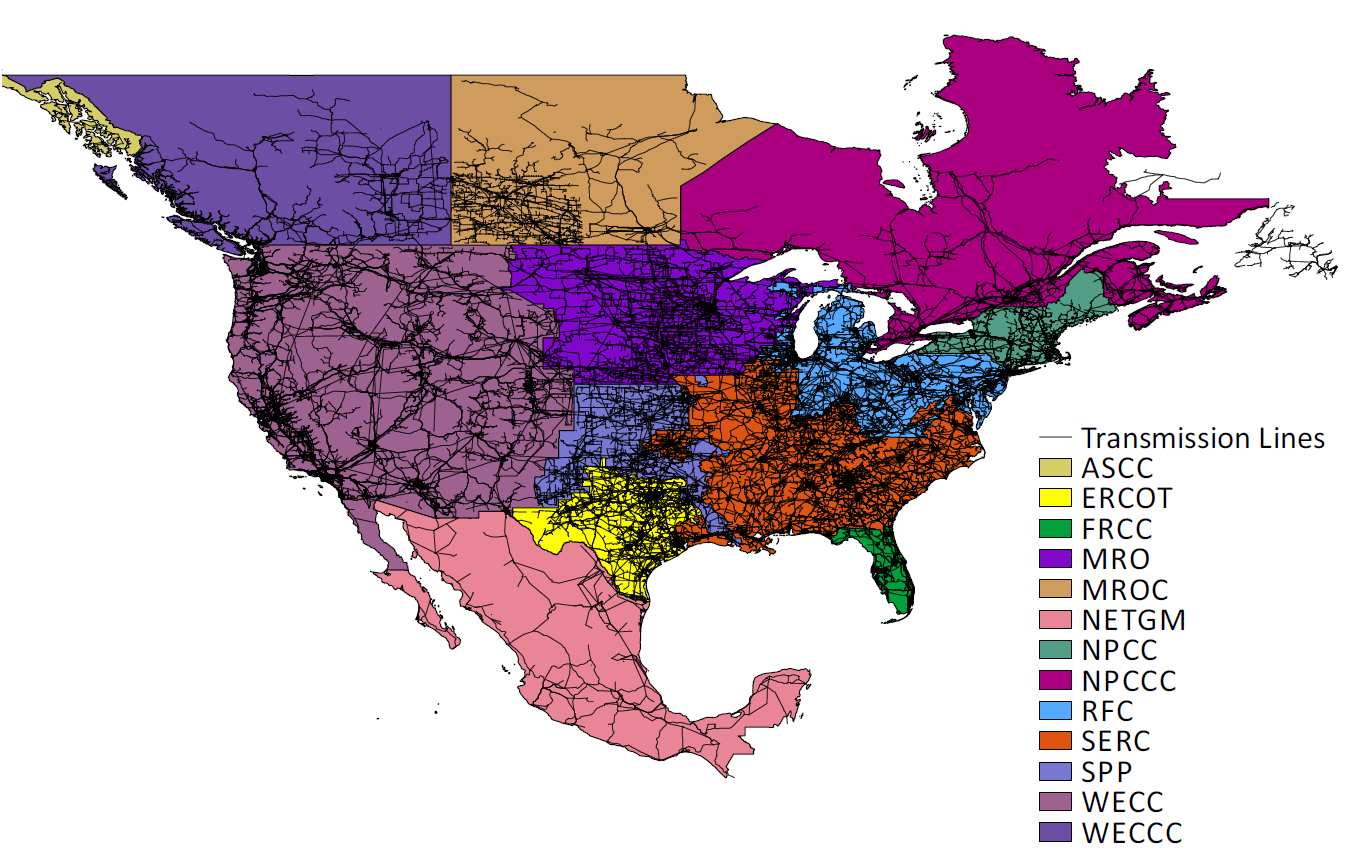}
\vspace*{-0.2cm}
\caption{The North American Electric Reliability Corporation (NERC) regional entities and the National Electricity Transmission Grid of Mexico (NETGM). Different reliability corporations/councils are marked with different colors.}
\label{fig:NERC}
\vspace*{0.2cm}
\end{figure}

\begin{figure}[t]
\centering
\vspace*{-0.2cm}
\includegraphics[scale=0.4]{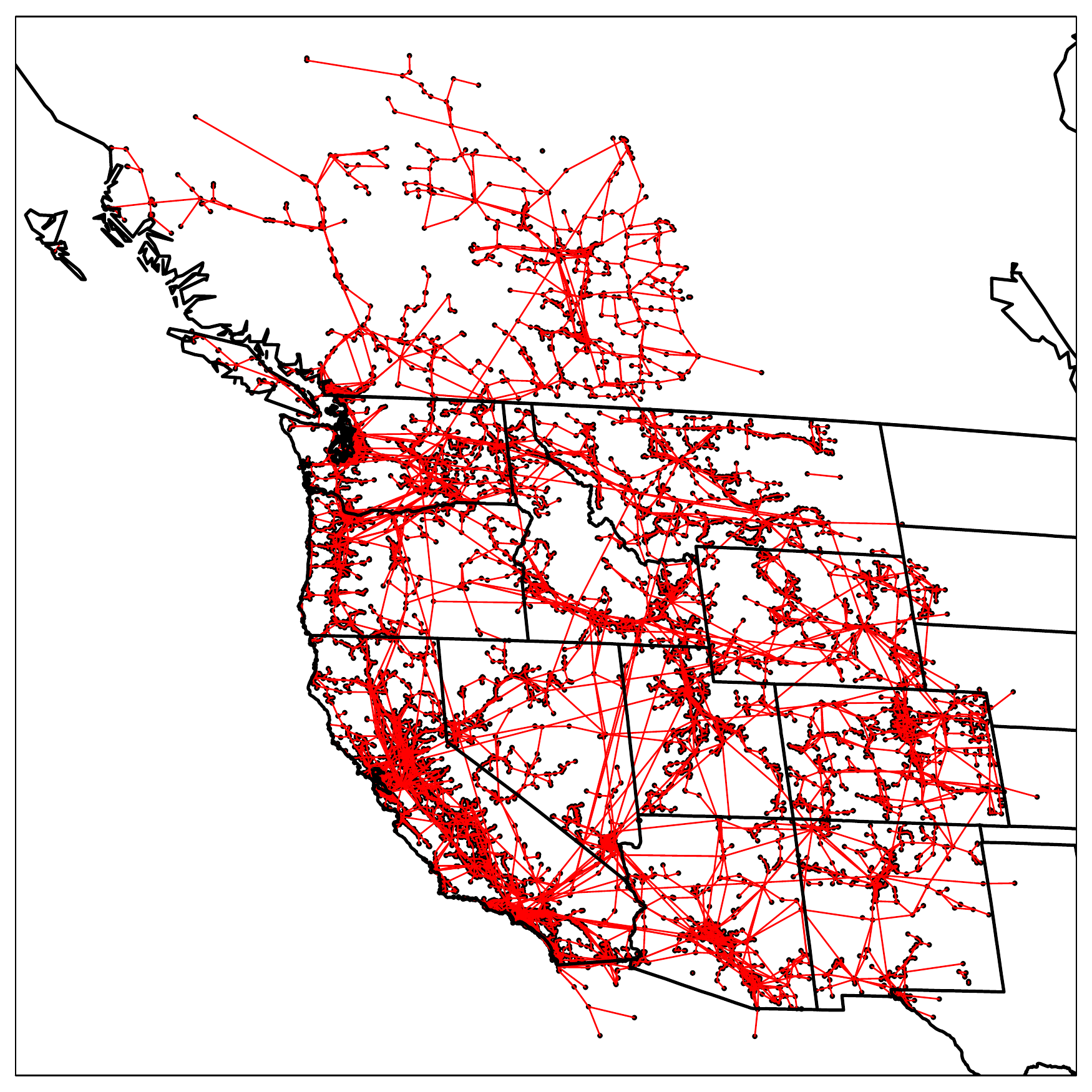}
\vspace*{-0.2cm}
\caption{The Western Interconnection (WI) power grid with 14,302 substations (nodes) and 18,769 lines (edges).}
\label{fig:WI}
\vspace*{0.2cm}
\end{figure}

To demonstrate the algorithm design and to evaluate its performance, we focus on the transmission networks of the North American and Mexican power grids (see NERC and NETGM in Fig.~\ref{fig:NERC}) using data that we obtained from the Platts Geographic Information System (GIS)~\cite{GIS}. We consider one of the two major interconnections -- the Western Interconnection (WI) (see Fig.~\ref{fig:WI}) which includes the Western Electricity Coordinating Council in the United States (WECC) and Canada (WECCC) (see Fig.~\ref{fig:NERC} for their coverage areas). Moreover, we consider two regional entities that operate under the Eastern Interconnection (EI) which is the other major interconnection -- the SERC Reliability Corporation (SERC), which is as large as the WI, and the Florida Reliability Coordinating Council (FRCC), which is much smaller than the WI.
To the best of our knowledge, \emph{this is the first time that the entire dataset of the North American and Mexican grids as well as those of SERC and FRCC are processed and analyzed}\footnote{Partial analysis of the WI dataset has been conducted before -- see Section \ref{sec:related}.}.

For the entire North American and Mexican grid as well as for WI, SERC, and FRCC, we consider four metrics that capture the networks' structural properties: average path length, clustering coefficient, degree distribution of the nodes, and the length distribution of the lines. The first three metrics are very common~\cite{watts1998collective,barabasi1999emergence,amaral2000classes,albert2004structural,crucitti2004topological,chassin2005evaluating,cotilla2012comparing}. However, to the best of our knowledge, \emph{the length distributions of the lines have not been thoroughly studied before}. These distributions are particularly important,
since the physical properties of a line (e.g., admittance and type) are directly correlated with its length~\cite{glover2011power}, and hence, the distributions directly impact the performance of various algorithms.

Motivated by the results of the structural properties' analysis, we present the \emph{Geographical Network Learner and Generator (GNLG) Algorithm for generating a network with similar properties to a given grid}.
First, using Gaussian Mixture Model (GMM), the algorithm estimates the density of the node positions and uses the obtained parameters to generate a set of nodes with a similar spatial distribution to these nodes (the algorithm uses the Bayesian Information Criterion (BIC) to find the best number of clusters for the GMM).
Then, the GNLG Algorithm uses two procedures, which are inspired by the historical evolution of power grids, to connect the generated nodes. Particularly, since the two main design considerations of the grid are connectivity and robustness, the algorithm obtains a spanning tree of the nodes to provide connectivity and then adds more edges to the network graph to increase its robustness. The addition of edges is tuned to create a synthetic network with properties that are similar to those of a given network.

To evaluate the performance of the GNLG Algorithm, we use it to generate networks similar to the WI, SERC, and FRCC. We show that by adapting a number of tunable parameters, the GNLG Algorithm can generate synthetic networks with similar structural and spatial properties to these power grid networks. Overall, we believe that by adapting the algorithm's tunable parameters, it is possible to generate synthetic networks similar to any given power grid network.

This paper is organized as follows. Section~\ref{sec:related} reviews related work. Section~\ref{sec:struc_prop} describes the dataset and the metrics, and presents the metrics for the different grids. Section~\ref{sec:gen_net} describes the GNLG Algorithm and Section~\ref{sec:eval} numerically evaluates its performance. We conclude and discuss future research directions in Section~\ref{sec:conclusion}.

\section{Related Work}\label{sec:related}
The structural properties of various power grids (e.g., in North America, some European countries, and Iran) were studied in~\cite{watts1998collective,rosas2007topological,sole2008robustness,
monfared2014topology,crucitti2004topological,danziger2015two}.
Most of these studies considered one or two properties (e.g., average degree, degree distribution, average path length, and clustering coefficient) and computed it in a given power grid. In some cases (e.g., \cite{watts1998collective,barabasi1999emergence,amaral2000classes,albert2004structural,crucitti2004topological,chassin2005evaluating,cotilla2012comparing}) a certain class of graphs was suggested as a good representative of a power grid network, based on one or two structural properties. For example, Watts and Strogatz ~\cite{watts1998collective} suggested the small-world graph as a good representative, based on the shortest path lengths between nodes and the clustering coefficient of the nodes. Barab\'{a}si and Albert~\cite{barabasi1999emergence} showed that scale-free graphs are better representatives based on the degree distribution. However, by comparing the WI with these models, Cotilla-Sanchez, et al.~\cite{cotilla2012comparing} showed that none of them can represent the WI properly.

More detailed models that are specifically tailored to the power grid characteristics  were proposed in~\cite{wang2010generating,schultz2014random} but they did not consider the nodes' \emph{spatial distribution} and the length distribution of the lines. The spatial distribution of the nodes is correlated with the length of the lines, and as mentioned above,
it is important to consider line lengths when designing a method for synthetic power grid generation.  While there are several models for generating spatial networks~\cite{marc2010spatial,manna2002modulated,xulvi2002evolving}, most of them were not designed to generate networks with properties similar to power grid networks. To the best of our knowledge, this paper is the first to consider the spatial distribution of the nodes in power grids and its importance in generating synthetic networks with similar structural properties.

\begin{figure*}[t]
\centering
\begin{subfigure}[b]{0.24\textwidth}
\vspace*{-0.2cm}
\includegraphics[scale=0.23]{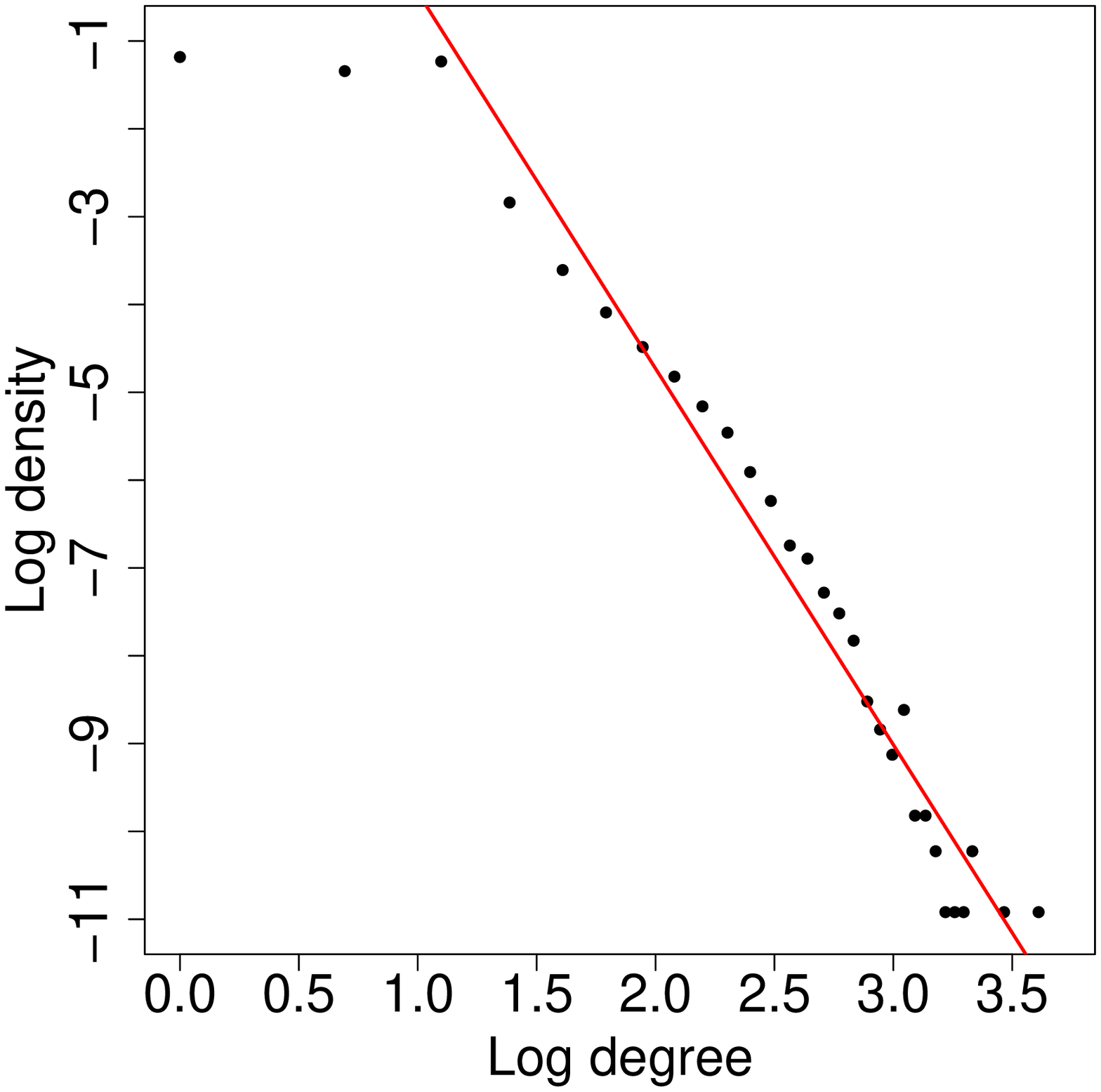}
\vspace*{-0.2cm}
\caption{NA\&M}
\end{subfigure}
\begin{subfigure}[b]{0.24\textwidth}
\vspace*{-0.2cm}
\includegraphics[scale=0.23]{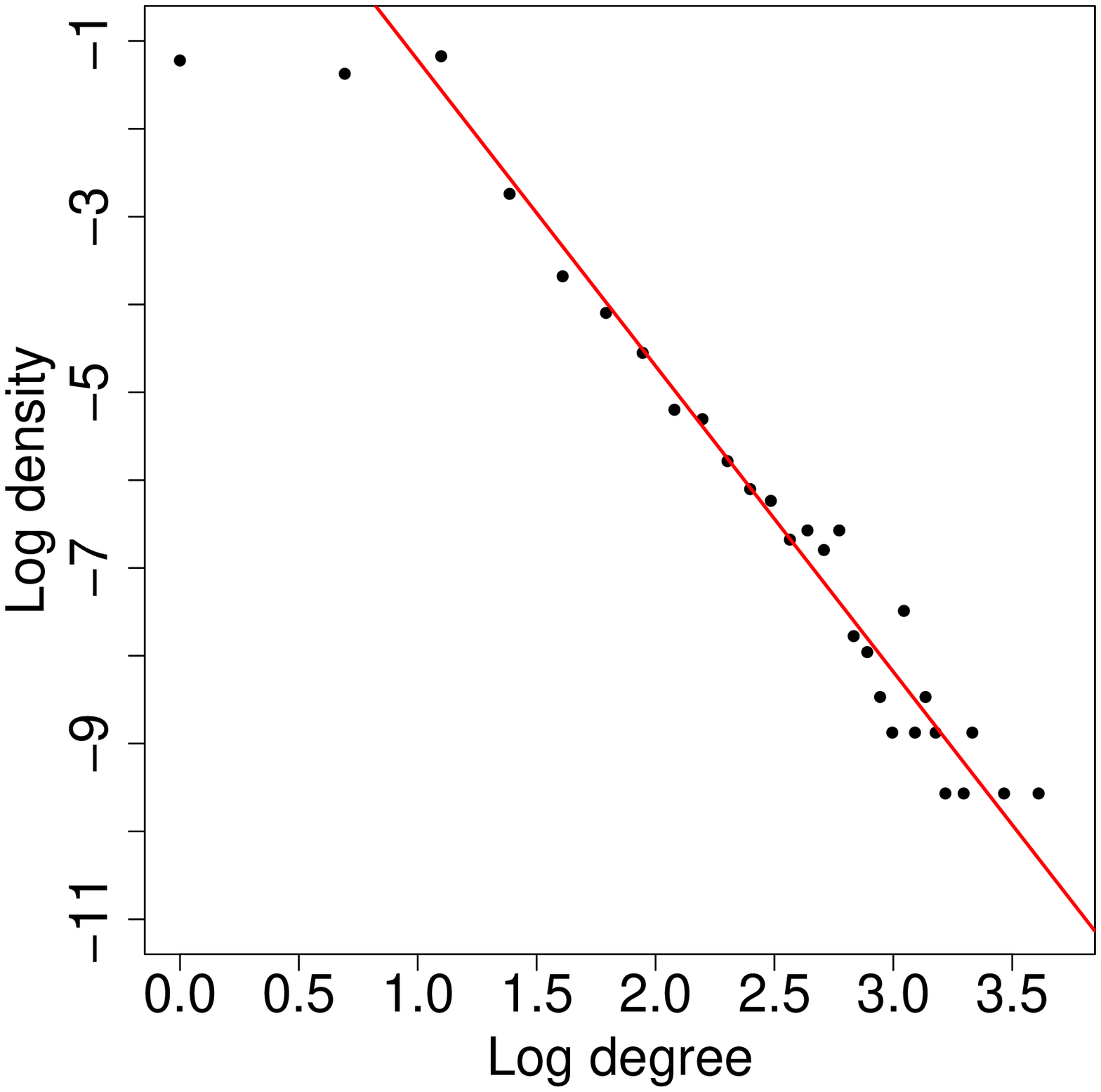}
\vspace*{-0.2cm}
\caption{WI}
\end{subfigure}
\begin{subfigure}[b]{0.24\textwidth}
\vspace*{-0.2cm}
\includegraphics[scale=0.23]{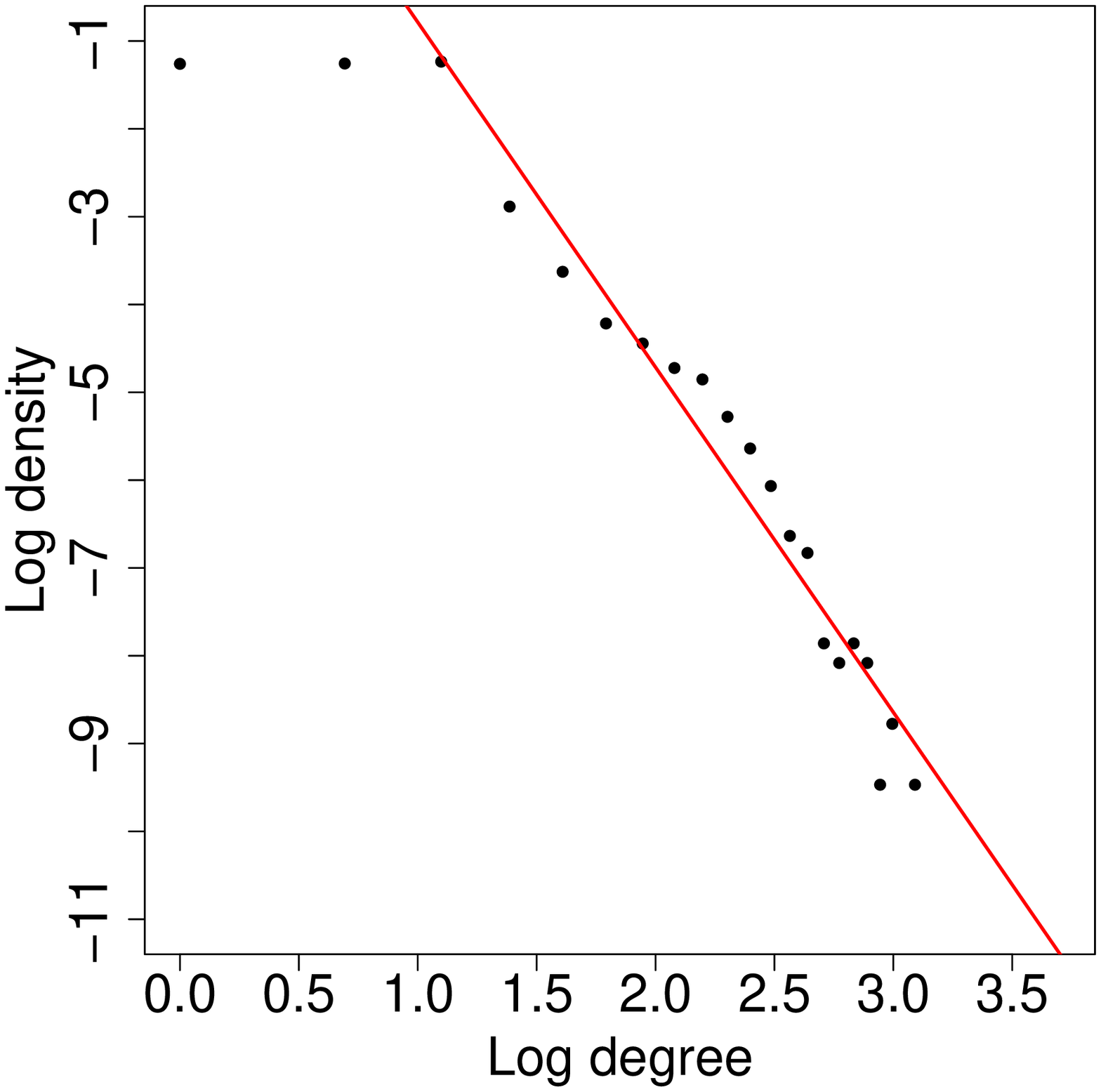}
\vspace*{-0.2cm}
\caption{SERC}
\end{subfigure}
\begin{subfigure}[b]{0.24\textwidth}
\vspace*{-0.2cm}
\includegraphics[scale=0.23]{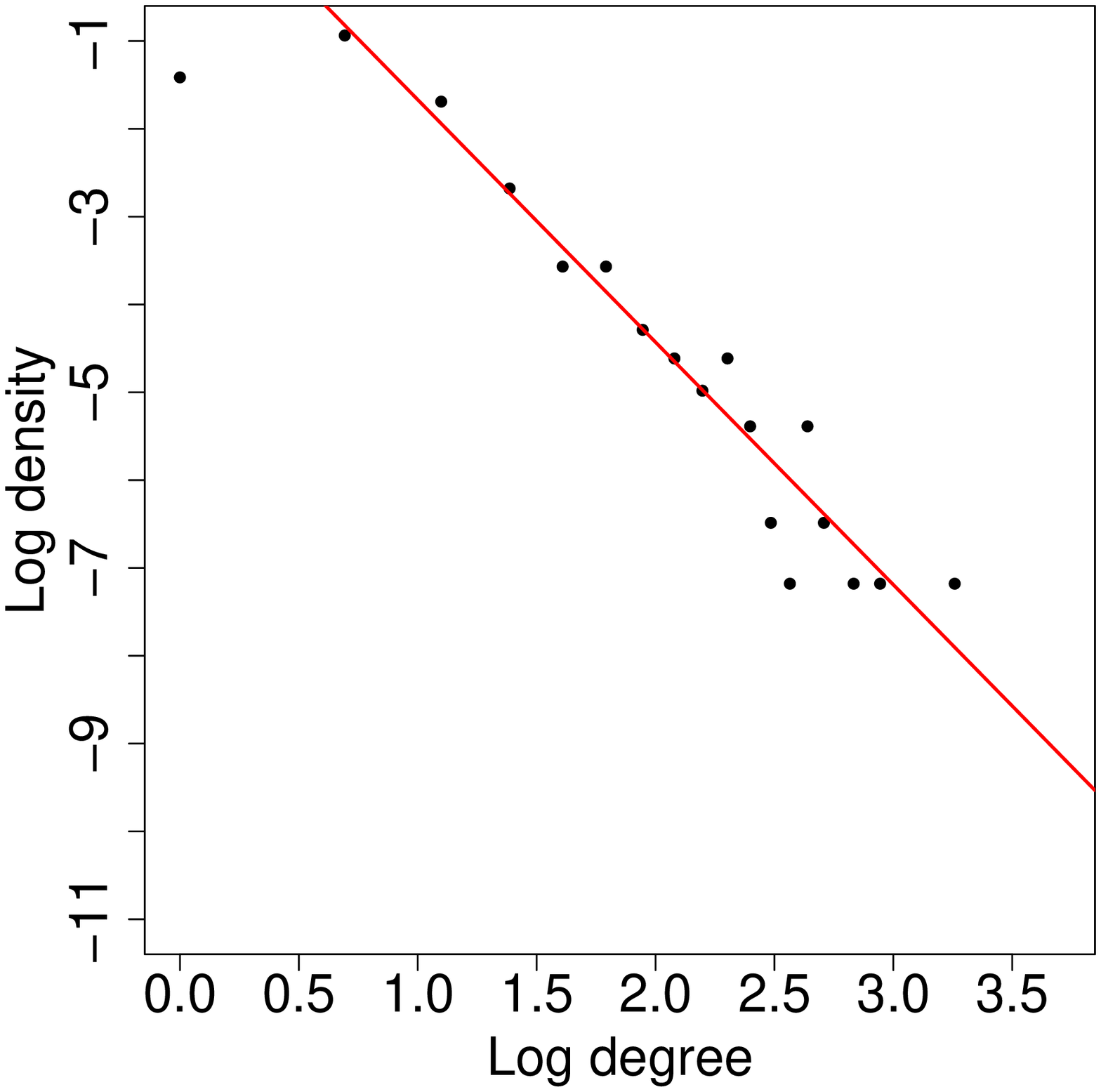}
\vspace*{-0.2cm}
\caption{FRCC}
\end{subfigure}
\caption{The degree distribution of the nodes in the NA\&M, WI, SERC, and FRCC grids (in
log-log scale). Linear regression lines with slopes $\zeta=-4.28$, $\zeta=-3.48$, $\zeta=-3.93$, and $\zeta=-2.76$, respectively, are fitted to
the tail distribution of the degrees.}
\label{fig:deg_dist_WI}
\vspace*{0.2cm}
\end{figure*}

\section{Preliminaries and Structural Properties}\label{sec:struc_prop}
In this section, we study the structural properties of the entire North American and Mexican grid (denoted by \mbox{NA\&M}) as well as of the WI, SERC, and FRCC grids.
We obtained the data from the Platts GIS~\cite{GIS} and conducted longitude-latitude to planar $(x,y)$ coordinate transformation, using the great-circle distance method. Since the files containing substations and files containing lines are not always consistent,
we extracted the coordinates of the substations from the end point coordinates of the lines.  We then used the geographical coordinates of the substations and the lines to construct the graphs with nodes and edges that represent substations and lines, respectively.  We used the map of reliability coorporations/councils boundaries to divide the graph into regional entities (as in Fig.~\ref{fig:NERC}).
To the best of our knowledge, beside~\cite{SmartGridComm11rep,Soltan2014Cascade} where an approximation of the WI graph was extracted from the Platts GIS dataset for simulations, it is the first time that this dataset is processed and analyzed.

\begin{table}[t]
\begin{center}
\caption{Summary of the structural properties of the NA\&M, WI, SERC, and FRCC grids.}
\begin{tabular}{|l|c|c|c|c|}
\hline
Network&NA\&M&WI&SERC&FRCC\\
\hline
Number of Nodes ($n$)&55,231& 14,302 &12,946 &1,312\\
\hline
Number of Edges ($m$)&70,088 &18,769 & 16,658 & 1,780\\
\hline
Average Path Length ($L$)&26.66 &17.33&19.71&11.68\\
\hline
Clustering Coefficient ($C$)&0.049 &0.049&0.049&0.075\\
\hline
Degree Distribution ($\zeta$)&-4.28 &-3.48&-3.93&-2.76\\
\hline
\end{tabular}\label{tb:summary_struct}
\end{center}
\end{table}

In addition to the number of the nodes and edges, we use four metrics for classifying the structural properties of these networks: \emph{average path length, clustering coefficient, degree distribution of the nodes, and length distribution of the lines.} Table~\ref{tb:summary_struct} includes these metrics for the NA\&M, WI, SERC, and FRCC grids.

\noindent\textbf{Notation.} We denote the WI, SERC, and FRCC  power grid transmission networks by graphs $G_{WI}$, $G_{SERC}$, and $G_{FRCC}$, respectively. For each network, $n$ and $m$ denote the number of the nodes and edges. $d_i$ denotes the degree of node $i$ and $\textbf{p}_i\in\mathbb{R}^2$ denotes its position. We define $\rho$ as the average Euclidean distance of a node from its $N$ nearest neighbors.
We use the prime symbol $(')$ to denote the values for a generated network (e.g., $G_{WI}'$ denotes the generated network).  All the logarithms in this paper are natural logarithms. All the geographical distances in this paper are Euclidean distances (i.e., $\|\textbf{p}_i-\textbf{p}_j\|_2$ is the distance between nodes $i$ and $j$).

\subsection{Average path length}
The average path length, denoted by $L$, is one of the common metrics used for classifying graphs. It is defined as the number of edges in the shortest path between two nodes, averaged over all pairs of vertices:
\begin{equation*}
L=\frac{1}{n(n-1)}\sum_{\substack{i\neq j\\i,j\in V}} \text{dist}(i,j),
\end{equation*}
where $\text{dist}(i,j)$ is the number of edges in the shortest path between nodes $i,j$. As can be seen in Table~\ref{tb:summary_struct}, the average path length in all the four networks is in $O(\log(n))$ which is very small and suggests that these networks have the small-world property.

\begin{figure*}[t]
\centering
\begin{subfigure}[b]{0.24\textwidth}
\vspace*{-0.2cm}
\includegraphics[scale=0.23]{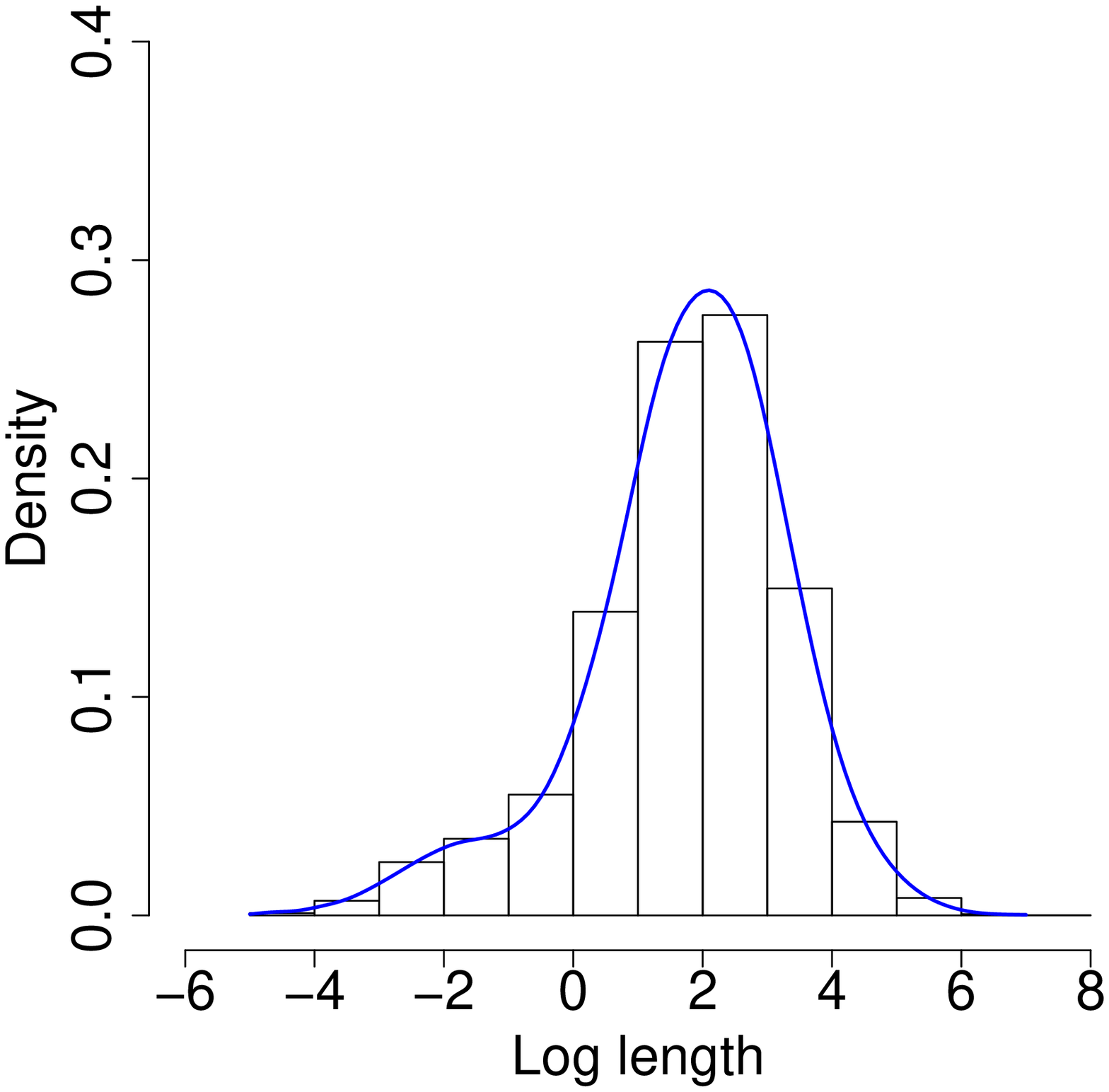}
\vspace*{-0.2cm}
\caption{NA\&M}
\end{subfigure}
\begin{subfigure}[b]{0.24\textwidth}
\vspace*{-0.2cm}
\includegraphics[scale=0.23]{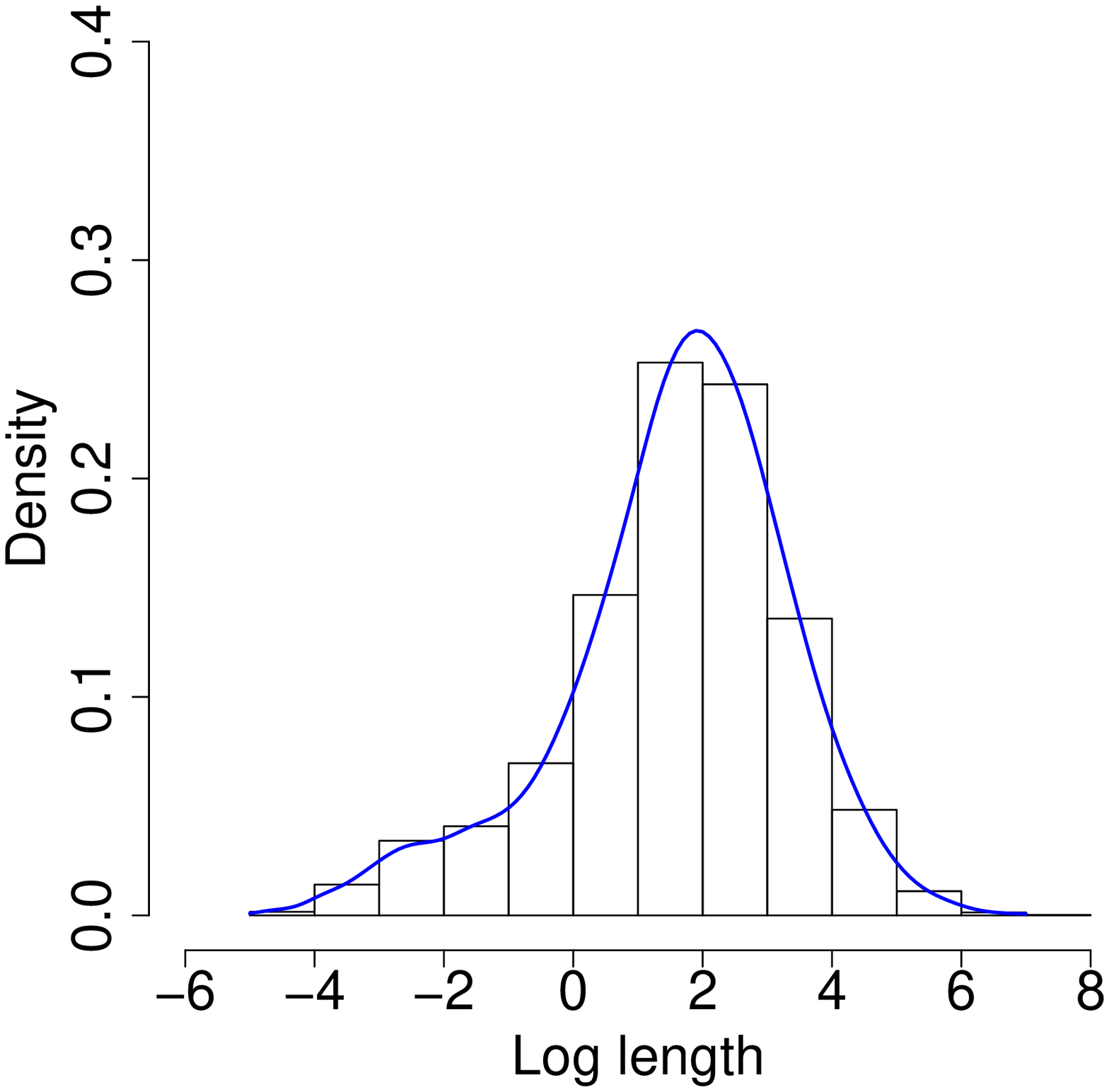}
\vspace*{-0.2cm}
\caption{WI}
\end{subfigure}
\begin{subfigure}[b]{0.24\textwidth}
\vspace*{-0.2cm}
\includegraphics[scale=0.23]{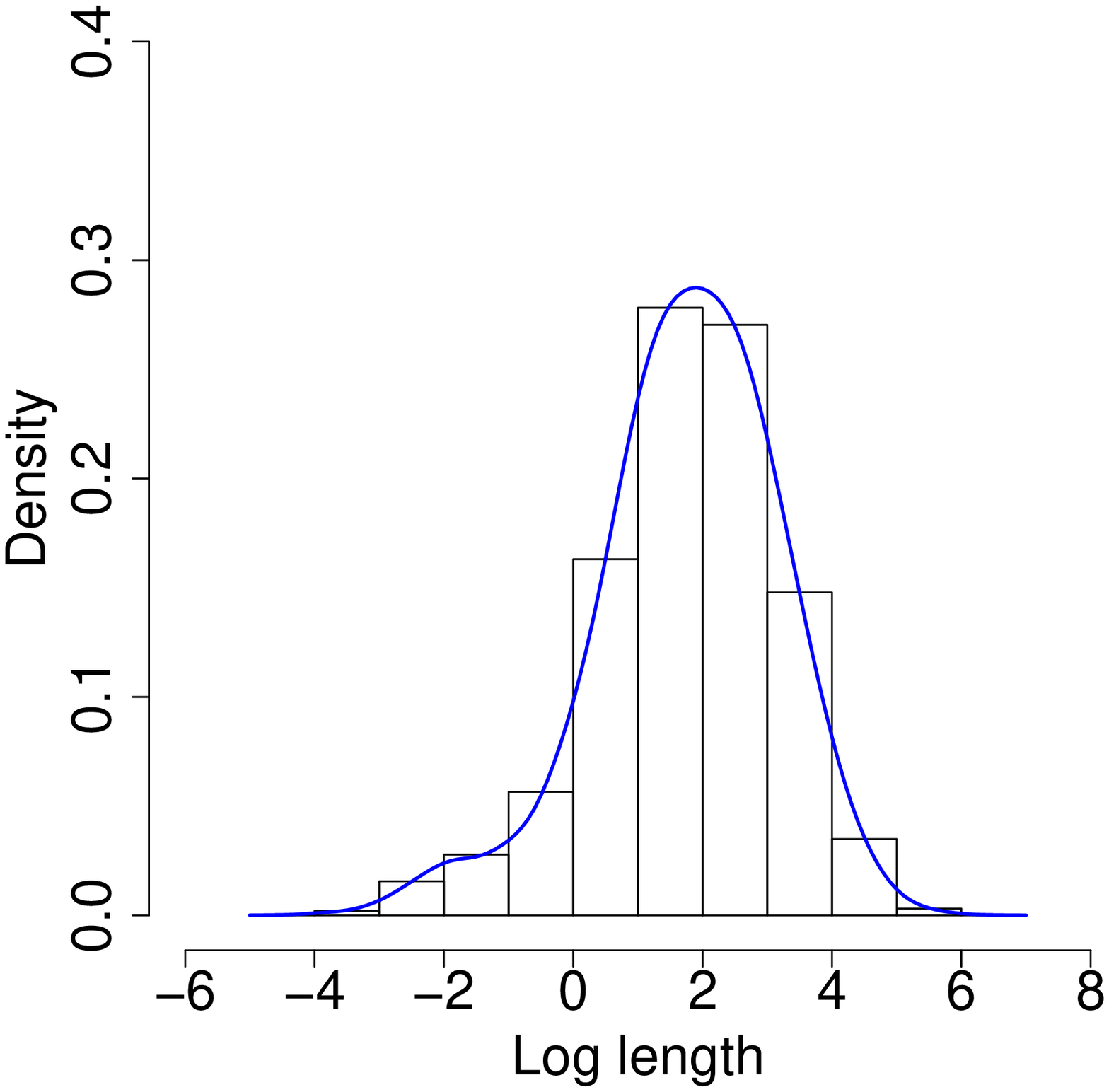}
\vspace*{-0.2cm}
\caption{SERC}
\end{subfigure}
\begin{subfigure}[b]{0.24\textwidth}
\vspace*{-0.2cm}
\includegraphics[scale=0.23]{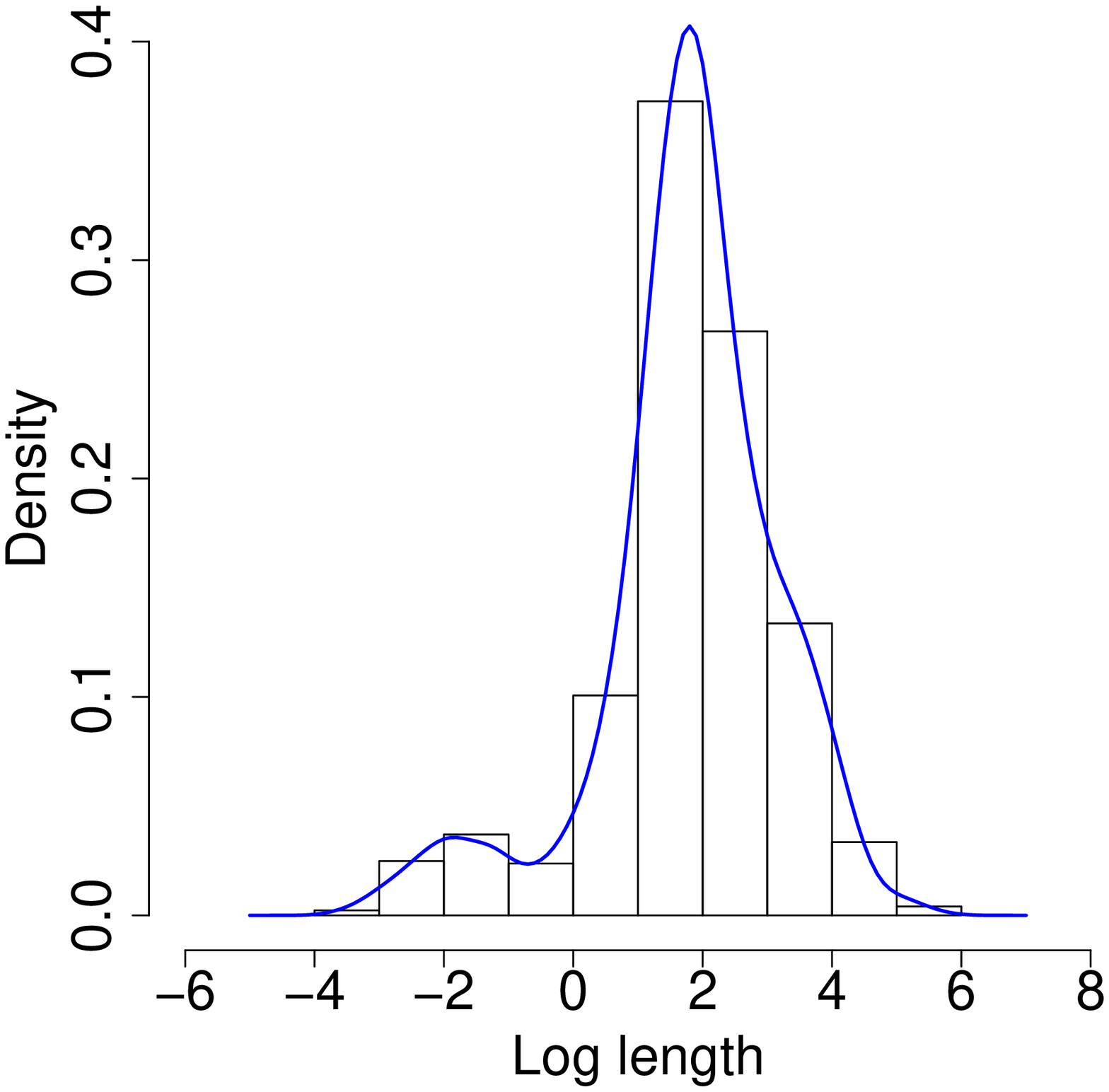}
\vspace*{-0.2cm}
\caption{FRCC}
\end{subfigure}
\caption{The distributions of the actual line lengths (in \emph{km}) in the NA\&M, WI, SERC, and FRCC grids (the lengths' statistics appear in Table~\ref{tb:summary_line_length}). Nonparametric distribution fits to the log length distributions are shown in blue.}
\label{fig:dist_lines_WI}
\vspace*{0.2cm}
\end{figure*}

\subsection{Clustering coefficient}
An important metric is the clustering coefficient, denoted by $C$ and defined as follows. For each node $i$, with degree $d_i$ at most $d_i(d_i-1)/2$ edges can exist
between its neighbors $N(i)$. Let $C_i$ denotes the fraction of these allowable edges that actually exist:
\begin{align*}
C_i &= \frac{|\{\{r,s\}|r,s\in N(i),\{r,s\}\in E\}|}{d_i(d_i-1)/2}.\\
\end{align*}
Then, averaging $C_i$ over all the nodes: $C = \sum_{i\in V} C_i / n$.
As can be seen in Table~\ref{tb:summary_struct}, the clustering coefficient for all the four networks is very small.

\subsection{Degree distribution of the nodes}
The degree distribution of the nodes is another important metric for classifying graphs (e.g., scale-free networks). Fig.~\ref{fig:deg_dist_WI} shows the degree distribution of the nodes in the NA\&M, WI, SERC, and FRCC grids in log-log scale. The degree one nodes in these networks usually correspond to power plants or small towns. These figures may suggest that the tail of the degree distribution follows a power-law distribution in all the three networks. However, following~\cite{clauset2009power} and since these networks are finite, we do not have enough statistical evidence to support the power-law hypothesis. Therefore, we only use the slope ($\zeta$) of the fitted linear regression line to the tail distribution for comparison purposes.

In Section~\ref{sec:eval}, we  use the Kolmogrov-Smirnov (KS) statistic~\cite{press2007numerical} to compare the degree distribution of the nodes in a given network and a generated network. If $P(x)$ and $Q(x)$ are two Cumulative Distribution Functions (CDFs), the KS statistic between these two is defined as follows:
\begin{equation*}
D_{KS} = \max_x |P(x)-Q(x)|.
\end{equation*}

\begin{figure}[t]
\centering
\vspace*{-0.2cm}
\includegraphics[scale=0.25]{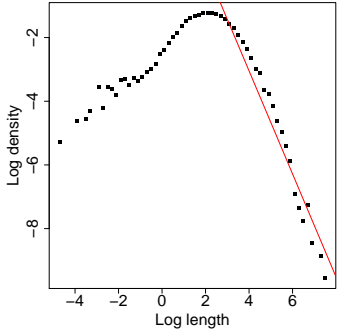}
\vspace*{-0.2cm}
\caption{The distribution of the actual line lengths (in \emph{km}) in the NA\&M grid in log-log scale. A linear regression line with slope $-1.61$ is fitted to the tail distribution of the lengths.}
\label{fig:dist_lines_NA}
\vspace*{0.2cm}
\end{figure}

\begin{table}[t]
\begin{center}
\caption{Statistics of the actual line lengths in the NA\&M, WI, SERC, and FRCC grids and of the corresponding straight lines (Euclidean distances) between substations in those grids (in $km$). The statistics of the straight lines are shown in the grey cells.}
\begin{tabular}{|l|c|c|c|c|}
\hline
Network&NA\&M&WI&SERC&FRCC\\
\hline
{\multirow{2}{*}{Mean}}& 15.46 &16.63 &13.29&12.82\\
 \cline{2-5} & \cellcolor[gray]{0.8} 14.30 &\cellcolor[gray]{0.8} 15.78 &\cellcolor[gray]{0.8} 11.39&\cellcolor[gray]{0.8} 9.95\\
\hline

{\multirow{2}{*}{Standard Deviation}}& 32.55 & 43.91 & 22.29&20.14\\

\cline{2-5}& \cellcolor[gray]{0.8}30.68 & \cellcolor[gray]{0.8}40.78 & \cellcolor[gray]{0.8}17.90&\cellcolor[gray]{0.8}15.6\\
\hline

{\multirow{2}{*}{Maximum}}& 1,714.82&1,714.82&795.44&282.74\\
\cline{2-5}& \cellcolor[gray]{0.8}1,380.35&\cellcolor[gray]{0.8}1,380.35&\cellcolor[gray]{0.8}409.92&\cellcolor[gray]{0.8}226.25\\
\hline
\end{tabular}\label{tb:summary_line_length}
\end{center}
\end{table}

\subsection{Length distribution of the lines}
As mentioned above, the length distribution of the lines is one of the important parameters that needs to be sustained in synthetic power grid generation. Fig.~\ref{fig:dist_lines_WI} shows the length distribution  of the lines in the NA\&M, WI, SERC, and FRCC grids. The length  distribution of the lines in the NA\&M grid in log-log scale is shown in Fig.~\ref{fig:dist_lines_NA}.\footnote{As can be seen in Figs.~\ref{fig:dist_lines_WI} and \ref{fig:dist_lines_NA}, there are some very short lines ($\approx 30 m$) in the considered networks. We checked the dataset to verify the credibility of these lines and did not find any issues (these lines are categorized as \emph{below 230kV} lines).} The lengths' statistics appear in Table~\ref{tb:summary_line_length}.

The line lengths in Figs.~\ref{fig:dist_lines_WI} and~\ref{fig:dist_lines_NA} are the \emph{actual lengths} of the power lines (these lines are not necessarily straight lines between two substation). To enable the comparison between the length distributions of the lines in the real and generated networks, in Section~\ref{sec:eval} we use the \emph{point-to-point Euclidean distances} to represent the line lengths in the real and the generated networks.  Table~\ref{tb:summary_line_length} includes the statistics regarding both the actual line lengths and the lengths of the straight lines between the substations, in order to demonstrate the differences between the metrics.

In Section~\ref{sec:eval}, we use Kullback-Leibler (KL) divergence to measure the similarity between the length distribution of the lines in a given network and a generated network.  The KL-divergence is a non-symmetric measure of the difference between two probability distribution functions $p$ and $q$. Specifically, the KL-divergence of $q$ from $p$, denoted $D_{KL}(p\|q)$, is a measure of the information lost when $q$ is used to approximate $p$:
 \begin{equation*}
 D_{KL}(p\|q)=\int_{-\infty}^{\infty} p(x)\ln\frac{p(x)}{q(x)} dx.
 \end{equation*}
 To estimate the KL-divergence between distributions, we use the {\ttfamily FNN} library in {\ttfamily R} which utilizes the method introduced in~\cite{boltz2009high} for estimating the KL-divergence between two distributions using their samples.

\section{Generating a Synthetic Network}\label{sec:gen_net}
In this section, we introduce the Geographical Network Learner and Generator (GNLG) Algorithm (Algorithm~\ref{alg:GSN}) for generating a synthetic network similar to a given network. The algorithm uses the Gaussian Mixture Model (GMM) for density estimation of the node positions and generates a set of nodes with similar spatial distribution to the nodes in a given network (the SDNG Procedure described in Subsection~\ref{sec:node_pos}). Then, it connects the nodes using two procedures whose design principles are inspired by historical evolution of the grids (the TWST and Reinforcement procedures described in Subsection~\ref{sec:node_con}).
The GNLG Algorithm can be applied to any network, where the important part is tuning the parameters to a given network.
In the following subsections, we describe the building blocks of the GNLG Algorithm and use the WI to demonstrate the algorithm design and operation. Then, in Section~\ref{sec:eval}, we evaluate the algorithm using the WI, SERC, and FRCC grids.

\begin{algorithm}[t]
\footnotesize
\caption{Geographical Network Learner and Generator (GNLG)}
\begin{trivlist}
\item\textbf{Input:} $G$, $\{\textbf{p}_i\}_{i=1}^n$, and parameters $\kappa,\alpha,\beta,\gamma>0$ and $N\in \mathbb{N}$.
\end{trivlist}
\vspace*{-3mm}
\begin{algorithmic}[1]
\STATE Generate a set of nodes with similar spatial distribution to the nodes \\in $G$ using the SDNG Procedure (Subsection~\ref{sec:node_pos}).
\STATE Connect the generated nodes using the TWST Procedure (Subsection~\ref{sec:node_con}).
\STATE Add more edges to the generated graph using the \\Reinforcement Procedure (Subsection~\ref{sec:node_con}).
\STATE \textbf{return} the generated graph $G'$.
\end{algorithmic}
\label{alg:GSN}
\end{algorithm}

\subsection{Node positions}\label{sec:node_pos}

We now introduce the Spatially Distributed Nodes Generator (SDNG) Procedure (Procedure~\ref{pro:SDNG}) for  generating a set of nodes with similar spatial distribution to the nodes in a given network. The node positions are correlated with the population and geographical properties (e.g.,  Fig.~\ref{fig:WI}). Thus, the nodes can be clustered into groups based on their geographical proximity. Mixture models and in particular Gaussian Mixture Models (GMM) are commonly used for clustering and density estimation~\cite{fraley2002model}. Hence, the SDNG Procedure uses the GMM for clustering the positions and uses BIC to find the best number of clusters ($c$). It obtains the mean and covariance matrix ($\mu_j, \Sigma_j$) of the points in clusters $j=1,\dots, c$  along with the categorical probability of the clusters $\pi=(\pi_1,\dots,\pi_{c})$. Then, it uses these parameters to generate $n$ nodes with similar spatial distribution as the nodes in a given network.

\begin{figure}[t]
\centering
\vspace*{-0.2cm}
\includegraphics[scale=0.4]{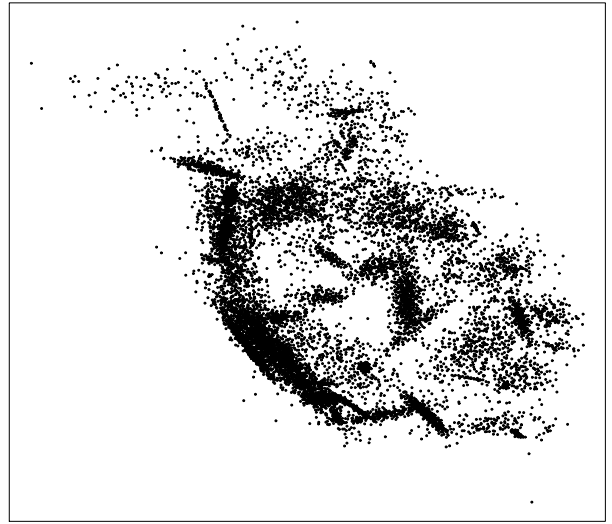}
\vspace*{-0.2cm}
\caption{A set of nodes, that were generated using the SDNG Procedure, with a similar spatial distribution to the nodes in the WI.}
\label{fig:WI_gen_nodes_55}
\vspace*{0.2cm}
\end{figure}

For implementing the SDNG Procedure, we used the {\ttfamily mclust} library in {\ttfamily R}~\cite{fraley2012mclust} to apply GMM to our dataset. This library uses the Expectation Maximization (EM) algorithm to fit a GMM and provides the Bayesian Information Criterion (BIC) for the selected number of clusters.
Clustering the nodes in the WI into 55 clusters results in the maximum BIC. Hence, the SDNG Algorithm clusters WI into $c=55$ clusters. As can be seen in Fig.~\ref{fig:WI_gen_nodes_55}, the distribution of the generated nodes appears very similar to the distribution of the nodes in the WI.

Notice that for a given network, step 1 in the Procedure should be executed only once. Then, having the fitted GMM parameters, the procedure can be used to generate several instances of  nodes with similar spatial distribution to the nodes in the given network. Hence, once the parameters are available, synthetic grids can be generated with no need to access the real grid data.

\setcounter{algocf}{0}
\begin{procedure}[t]
\footnotesize
\caption{Spatially Distributed Nodes Generator () (SDNG)}
\begin{trivlist}
\item\textbf{Input:} $G$, $\{\textbf{p}_i\}_{i=1}^n$.
\end{trivlist}
\vspace*{-3mm}
\begin{algorithmic}[1]
\STATE Fit a GMM model to $\{\textbf{p}_i\}_{i=1}^n$ to cluster them into $c$ clusters that maximizes the BIC.
\STATE For all $i=1,\dots,n$ sample $z_i$ from the categorical probability distribution $\pi$ obtained from GMM.
\STATE For all $i$ sample $\textbf{p}'_i$ from the probability distribution $\mathcal{N}(\mu_{z_i},\Sigma_{z_i})$ obtained from GMM.
\STATE \textbf{return} $\{\textbf{p}_i'\}_{i=1}^n$.
\end{algorithmic}
\label{pro:SDNG}
\end{procedure}

\begin{figure}[t]
\centering
\vspace*{-0.2cm}
\includegraphics[scale=0.2]{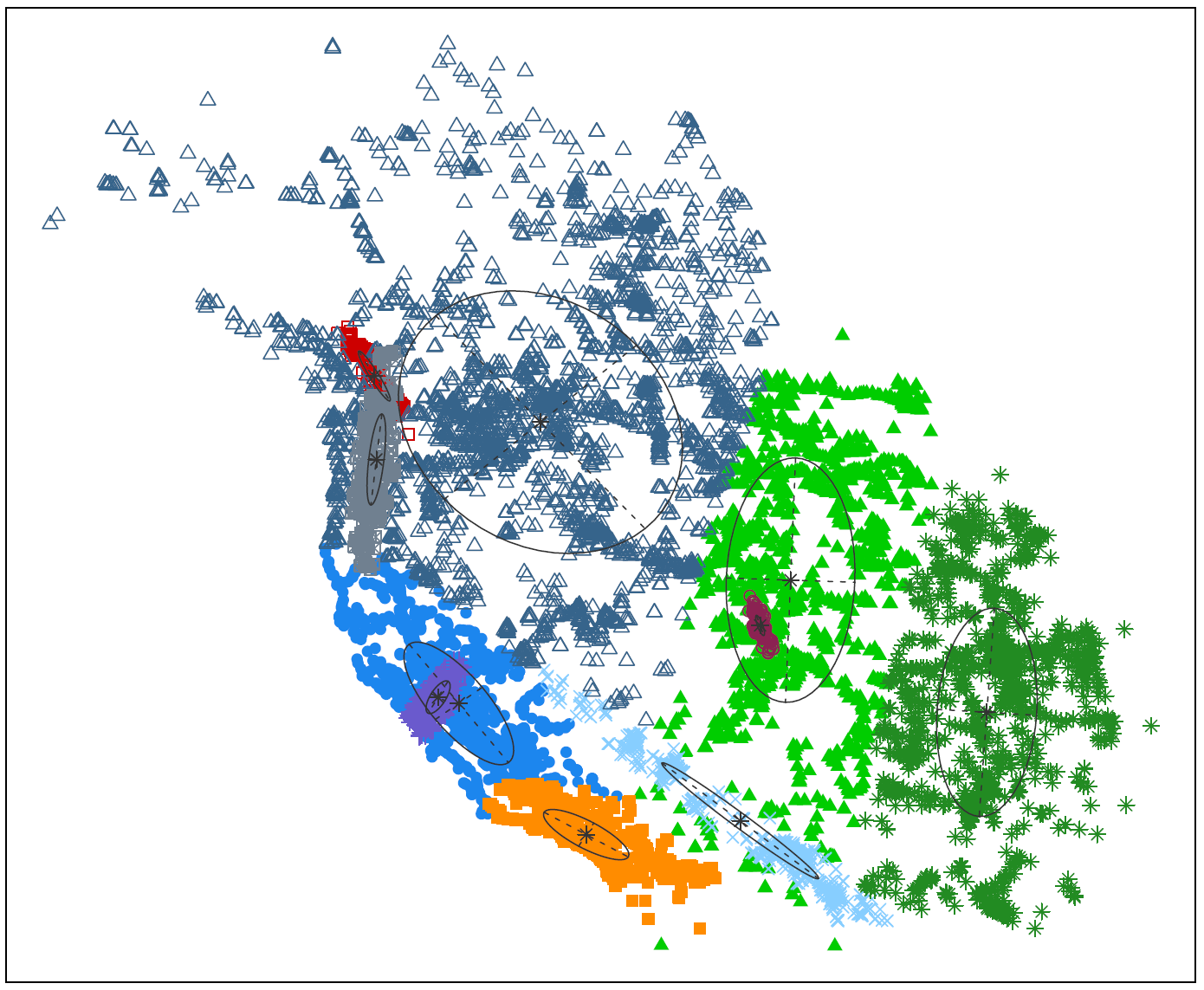}
\vspace*{-0.2cm}
\caption{An example of clustering the nodes in the WI into 10 clusters using GMM.}
\label{fig:WI_clust_10}
\vspace*{0.2cm}
\end{figure}

\subsection{Connections between the nodes}\label{sec:node_con}

We introduce two procedures (steps 2 and 3 in the GNLG Algorithm)  for connecting the generated nodes. Their design is inspired by the historical evolution of power grids.
The two main design consideration of the grid are (i) connectivity and (ii) robustness. Therefore, we first present the Tunable Weight Spanning Tree (TWST) Procedure for finding a spanning tree and to ensure connectivity. We then describe the Reinforcement Procedure for adding more edges and ensuring the network robustness as well as for tuning the structural properties of the synthetic network to resemble those of a given network.

\subsubsection{Connectivity}

\begin{figure}[t]
\centering
\begin{subfigure}[b]{0.24\textwidth}
\vspace*{-0.2cm}
\includegraphics[scale=0.23]{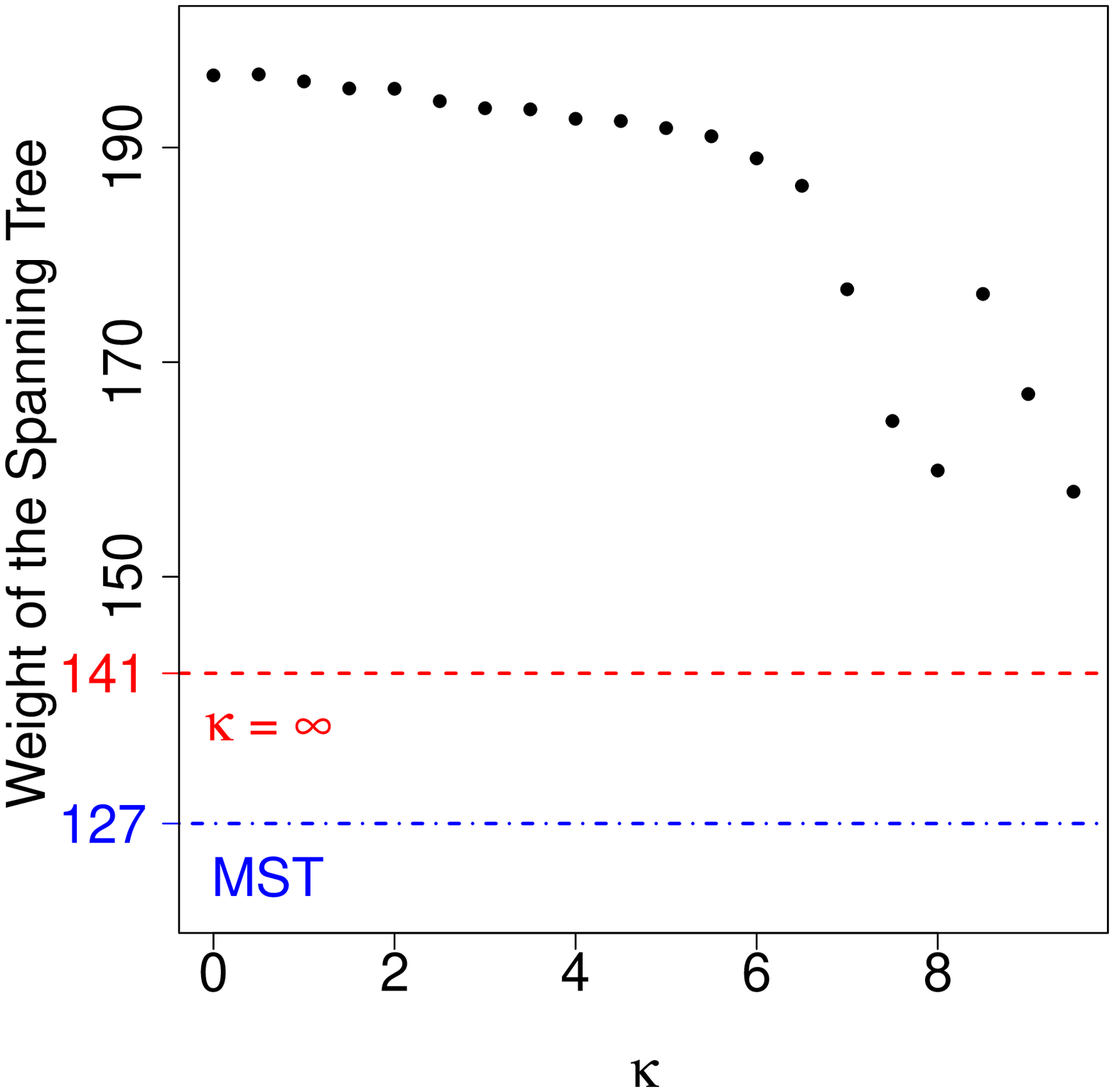}
\vspace*{-0.2cm}
\caption{}
\label{fig:kappa_span_weight}
\vspace*{0.2cm}
\end{subfigure}
\begin{subfigure}[b]{0.24\textwidth}
\vspace*{-0.2cm}
\includegraphics[scale=0.23]{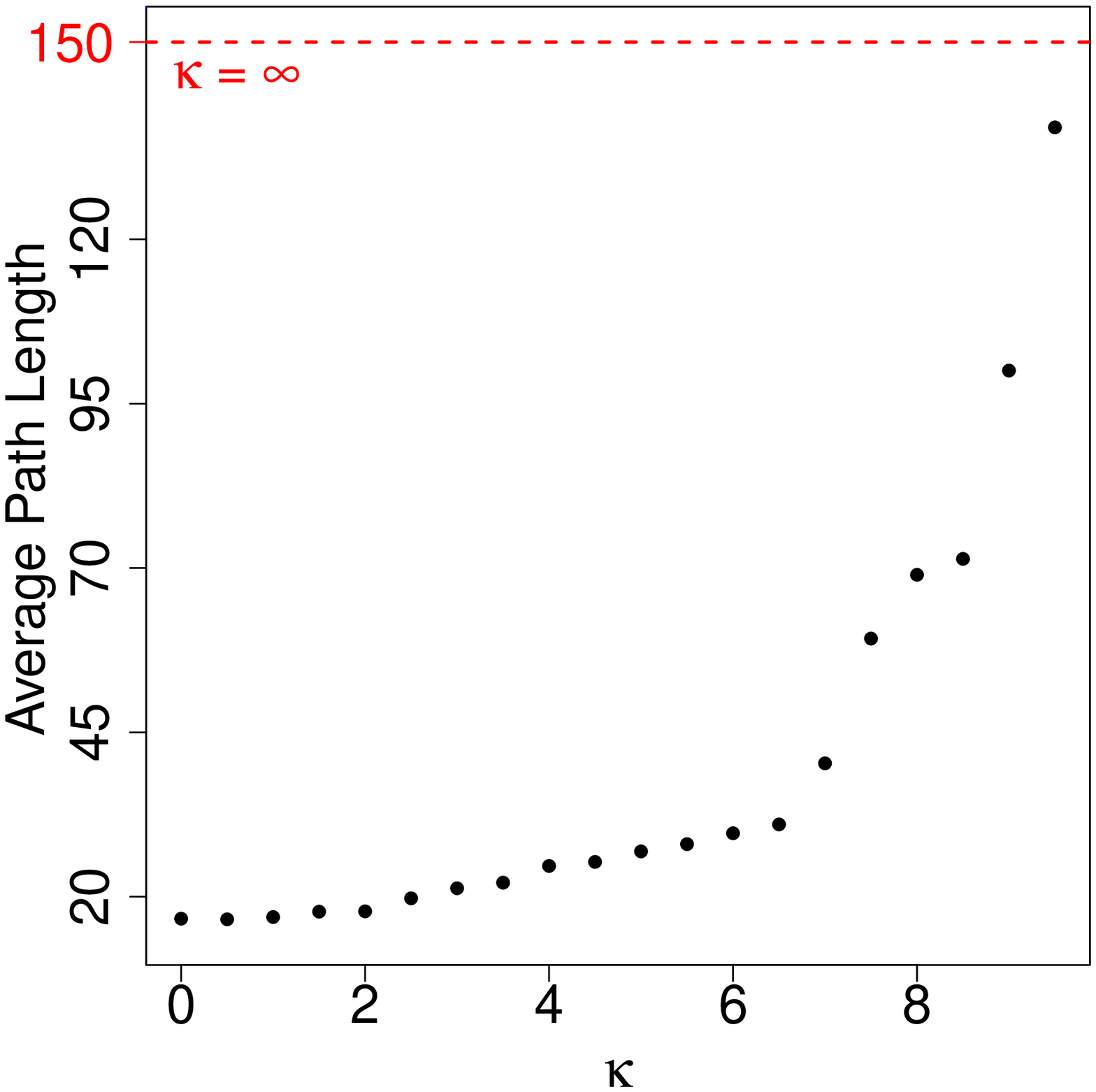}
\vspace*{-0.2cm}
\caption{}
\label{fig:kappa_ave_path}
\vspace*{0.2cm}
\end{subfigure}
\caption{(a) The weight of the spanning tree (in $10^3$\emph{km}) obtained by the TWST Procedure on the nodes shown in Fig.~\ref{fig:WI_gen_nodes_55} vs.\ $\kappa$. Each point is the average over 10 generated trees. The blue dash-dot line shows the weight of the MST and the red dashed line shows the weight of the obtained spanning tree for $\kappa=\infty$. (b) The average path length in the spanning tree obtained by the TWST Procedure on the nodes shown in Fig.~\ref{fig:WI_gen_nodes_55} vs.\ $\kappa$. Each point is the average over 10 generated trees. The average path length in a specific MST (an MST may not be unique) is 520. The red dashed line shows the average path length in the obtained spanning tree for $\kappa=\infty$.}
\label{fig:kappa}
\end{figure}

In order for the power grid to operate, the substations (nodes) should be connected. Due to construction costs, in the real world new substations are usually connected to the nearest substation in the existing grid. Since the power grids have evolved gradually and locally, they do not necessarily contain the Minimum weight Spanning Tree (MST) of the nodes in the plane (the weight of a spanning tree $T=(V_T,E_T)$ is the sum of the edge lengths in $T$:
 $W_T = \sum_{\{i,j\}\in E_T} \|\textbf{p}'_i-\textbf{p}'_j\|$).
 Hence, we do not focus on finding the MST. Instead, we present the TWST Procedure (Procedure~\ref{pro:span}), which imitates the the gradual grid evolution. It is a low complexity procedure for finding a spanning tree with a tunable weight.

The procedure uses the average node location, denoted by: $\bar{\textbf{p}}'=\sum_i \textbf{p}_i'/n$. It first orders the nodes in $n$ rounds (see step~\ref{step:for}) to obtain a permutation of indices $\sigma:\{1,2,\dots,n\}\rightarrow \{1,2,\dots,n\}$. At round $i$, it samples a node $j$ from the nodes that were not already sampled with probability proportional to $\|\textbf{p}_j'-\bar{\textbf{p}}'\|^{-\kappa}$, where $\kappa$ is a parameter. It then sets $\sigma(i)\leftarrow j$. In step~\ref{step:for2} it connects each node $\sigma(i)$ to its nearest neighbor $\sigma(j^*)$ such that $j^*<i$.

The procedure results in a tree whose weight highly depends on the ordering of the nodes, and thereby on $\kappa$. Moreover, there is a specific ordering of the nodes such that the procedure provides the MST (the nodes should be ordered according to their appearance in Prim's Algorithm~\cite{cormen2009introduction} for finding the MST).
Specifically, $\kappa$ determines the difference between the obtained spanning tree and the MST. Fig.~\ref{fig:kappa}(a) shows the relationship between the weight of the obtained tree and $\kappa$. When $\kappa=0$, the nodes are ordered randomly and the weight of the obtained spanning tree significantly differs from the MST's weight. However, As $\kappa$ increases the weight of the spanning tree decreases. When $\kappa$ is very large, the nodes are ordered based on their distance from the average location, and therefore, the obtained spanning tree's weight is close to the MST's (shown by the blue dash-dot line).

Fig.~\ref{fig:kappa}(b) shows the relationship between $\kappa$ and the average path length in the obtained tree. As $\kappa$ increases, the average path length increases. For large $\kappa$, this increase is more significant. Moreover, the average path length in an MST (520) is significantly larger than in trees obtained by the TWST Procedure. Overall, Figs.~\ref{fig:kappa}(a),(b) suggest that selecting a relatively small $\kappa$  results in a spanning tree with smaller average path length than the MST and with a reasonable total weight.  We show in Section~\ref{sec:eval} that for generating a network similar to the WI, $\kappa=2.5$ is a relatively good choice.

\begin{procedure}[t]
\footnotesize
\caption{Tunable Weight Spanning Tree() (TWST)}
\begin{trivlist}
\item\textbf{Input:} $n, \{\textbf{p}_i'\}_{i=1}^n$, and parameter $\kappa$.
\end{trivlist}
\vspace*{-3mm}
\begin{algorithmic}[1]
\STATE $A=\{1,\dots,n\}$, $\sigma$ is an empty array of size $n$.
\FOR {$i=1\dots,n$} \label{step:for}
\STATE Sample a node from $A$ such that the probability of sampling \\node $j$ is $\frac{\|\textbf{p}_j'-\bar{\textbf{p}}'\|^{-\kappa}}{\sum_{a\in A}\|\textbf{p}_a'-\bar{\textbf{p}}'\|^{-\kappa}}$.
\STATE $\sigma(i)$ $\leftarrow$ $j$, $A\leftarrow A\backslash\{j\}$.
\ENDFOR
\FOR {$i=2,\dots,n$}\label{step:for2}
\STATE Connect node $\sigma(i)$ to node $\sigma(j^*)$ such that $j^* = \text{argmin}_{j<i} \|\textbf{p}_{\sigma(i)}'-\textbf{p}_{\sigma(j)}'\|$.
\ENDFOR
\end{algorithmic}
\label{pro:span}
\end{procedure}

\subsubsection{Robustness}

\begin{figure}[t]
\centering
\vspace*{-0.2cm}
\includegraphics[scale=0.3]{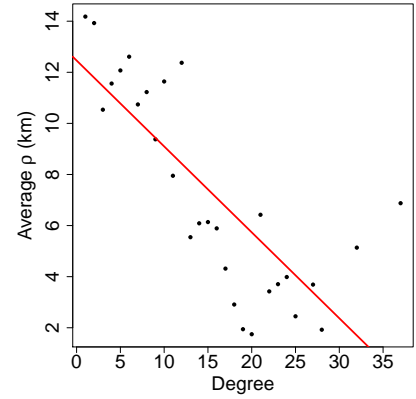}
\vspace*{-0.2cm}
\caption{The relationship between the degree of a node and its average $\rho$ with $N=10$, for the nodes in the WI
%$G_{WI}$
(the red line is the linear regression fit to the data points).}
\label{fig:rho_degree}
\vspace*{0.2cm}
\end{figure}

\begin{procedure}[t]
\footnotesize
\caption{Reinforcement()}
\begin{trivlist}
\item\textbf{Input:} $n, m, \{\textbf{p}'_i\}_{i=1}^n$, and parameters $\alpha,\beta,\gamma,\eta>0$, $N\in \mathbb{N}$.
\end{trivlist}
\vspace*{-3mm}
\begin{algorithmic}[1]
\STATE For each node $i$, compute $\rho_i$ (the average distance of node $i$ from \\its $N$ nearest neighbors).
\FOR {$count=1$ to $m-n+1$}
\STATE \textbf{if} \emph{large network}: From all nodes with degree less than 3,\\ sample node $i$  with probability $\propto \rho_i'^{-\alpha}$.
\STATE \textbf{if} \emph{small network}: Sample node $i$  with probability $\propto d_i'^{-\eta}\rho_i'^{-\alpha}$.
\STATE Connect node $i$ to node $j$ sampled from all other nodes with probability $\propto \|\textbf{p}'_i-\textbf{p}'_j\|^{-\beta} d_j'^{\gamma}$.
\ENDFOR
\end{algorithmic}
\label{pro:add_edge}
\end{procedure}
We present the Reinforcement Procedure whose objective is to increase the robustness of the generated network and adjust its properties (e.g., $L$ and $C$) to resemble those of a given network. The procedure is based on three observations: (i) the degree distributions of power grids are very similar to those of scale-free networks, but grids have less degree 1 and 2 nodes and do not have very high degree nodes (e.g., Fig.~\ref{fig:deg_dist_WI}), (ii) it is inefficient and unsafe for the power grids to include very long lines (e.g., Figs.~\ref{fig:dist_lines_WI} and \ref{fig:dist_lines_NA}), and (iii) nodes in denser areas are more likely to have higher degrees. The last observation is demonstrated by Fig.~\ref{fig:rho_degree} where as  the degree increases, the $\rho$ decreases\footnote{Recall that $\rho$ is the average Euclidean distance of a node from its $N$ nearest neighbors.} (i.e., the density around a node increases).

The Reinforcement Procedure aims to create a network whose properties are similar to those observed above. Hence, it repeats the following steps $m-n+1$ times: (1) selects a low degree node in a dense area (observations (i) and (iii)), and (2) connects it to a high degree node (as in the preferential attachment model~\cite{barabasi1999emergence}) which is also nearby (distance was not considered in~\cite{barabasi1999emergence}) (observations (i) and (ii)).

\emph{To select a low degree node in a dense area},
the Reinforcement Procedure samples a node $i$ with probability $\propto d_i ^{-\eta}\rho_i^{-\alpha}$. However, as can be seen in Fig.~\ref{fig:deg_dist_WI}, the distribution of the degree 1 and 2 nodes is almost equal in the WI and SERC grids.
Hence, for large networks, the procedure only considers degree 1 and 2 nodes and select a node among them with probability $\propto \rho_i^{-\alpha}$. $\alpha$ and $\eta$ are the tunable parameters.

\begin{figure}[t]
\centering
\vspace*{-0.2cm}
\includegraphics[scale=0.4]{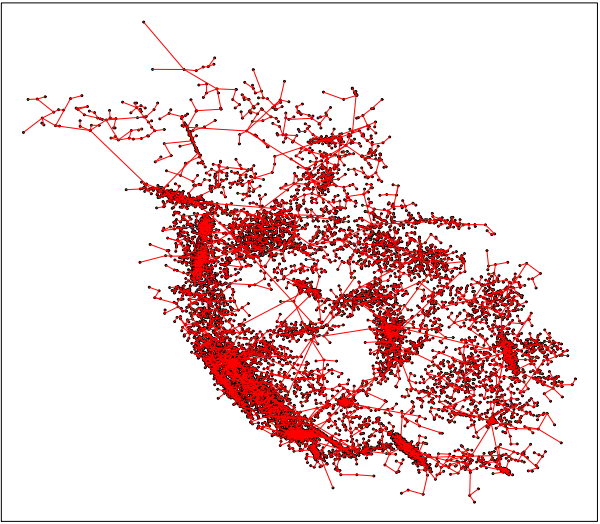}
\vspace*{-0.2cm}
\caption{A network with 14,302 nodes and 18,769 edges generated based on the WI grid using the GNLG Algorithm with $\kappa=2.5$, $\alpha = 1, \beta = 3.2, \gamma = 2.5,$ and $N=10$.}
\label{fig:Generated_WI}
\vspace*{0.2cm}
\end{figure}

\emph{To connect the node sampled in the previous step to a high degree but nearby node}, in the second step, the Reinforcement Procedure connects node $i$ to node $j$ sampled from all other nodes with probability $\propto \|\textbf{p}'_i-\textbf{p}'_j\|^{-\beta} d_j'^{\gamma}$.
This implies that node $i$ preferentially connects to a high-degree node, unless the high-degree node is too far in which case it is desirable to connect to a low-degree but nearby node. This is very similar to the model introduced in~\cite{manna2002modulated,xulvi2002evolving}. However, here we only use these probabilities for sampling and do not use them for connecting every pair of nodes.

We note that $\beta$ determines the length distribution of the new lines and $\gamma$ determines the likelihood of the existence of high degree nodes. If $\beta$ is large compared to $\gamma$, then new edges connect nearby nodes, thereby resulting in a large clustering coefficient and  a large average path length. If $\gamma$ is large compared to $\beta$, then new edges connect nodes to high degree nodes regardless of their distance, thereby resulting in very high degree nodes and long edges. Hence, there should be a balance between the $\beta$ and $\gamma$ values. We show in Section~\ref{sec:eval} that for generating a network similar to the WI, $\beta=3.2$ and $\gamma=2.5$ are relatively good choices.

\section{Evaluation}\label{sec:eval}
In this section, we use the GNLG Algorithm to generate networks similar to the WI, SERC, and FRCC grids. We evaluate the structural properties of the obtained networks and show that they have similar properties to the real networks.
\subsection{WI}
As mentioned in Section~\ref{sec:node_con}, the parameters $\kappa,\alpha,\beta,\gamma,N$ can be used to tune the structural properties of the obtained network. Therefore, we conducted several numerical experiments in which the parameters were adapted and the structural properties were evaluated. We observed empirically that the following parameters values provide a network with similar properties to the WI: $\kappa=2.5$, $\alpha = 1, \beta = 3.2, \gamma = 2.5,$ and $N=10$. Moreover, as mentioned in Section~\ref{sec:node_pos}, BIC was used to determine the number of clusters ($c=55$).

The nodes generated by the SDNG Procedure were shown in Fig.~\ref{fig:WI_gen_nodes_55}. The network obtained by the GNLG Algorithm  appears in Fig.~\ref{fig:Generated_WI} and visually resembles the WI. To study the structural similarity between the obtained network $G_{WI}'$ and the $G_{WI}$, we evaluated $G_{WI}'$ based on the metrics described in Section~\ref{sec:struc_prop}. The clustering coefficient and the average path length of $G_{WI}'$ are $C'=0.052$ and $L'=17.40$, respectively, and are very close to those of  $G_{WI}$ ($C=0.049$ and $L=17.33$).

\begin{figure}[t]
\centering
\begin{subfigure}[b]{0.24\textwidth}
\vspace*{-0.2cm}
\includegraphics[scale=0.25]{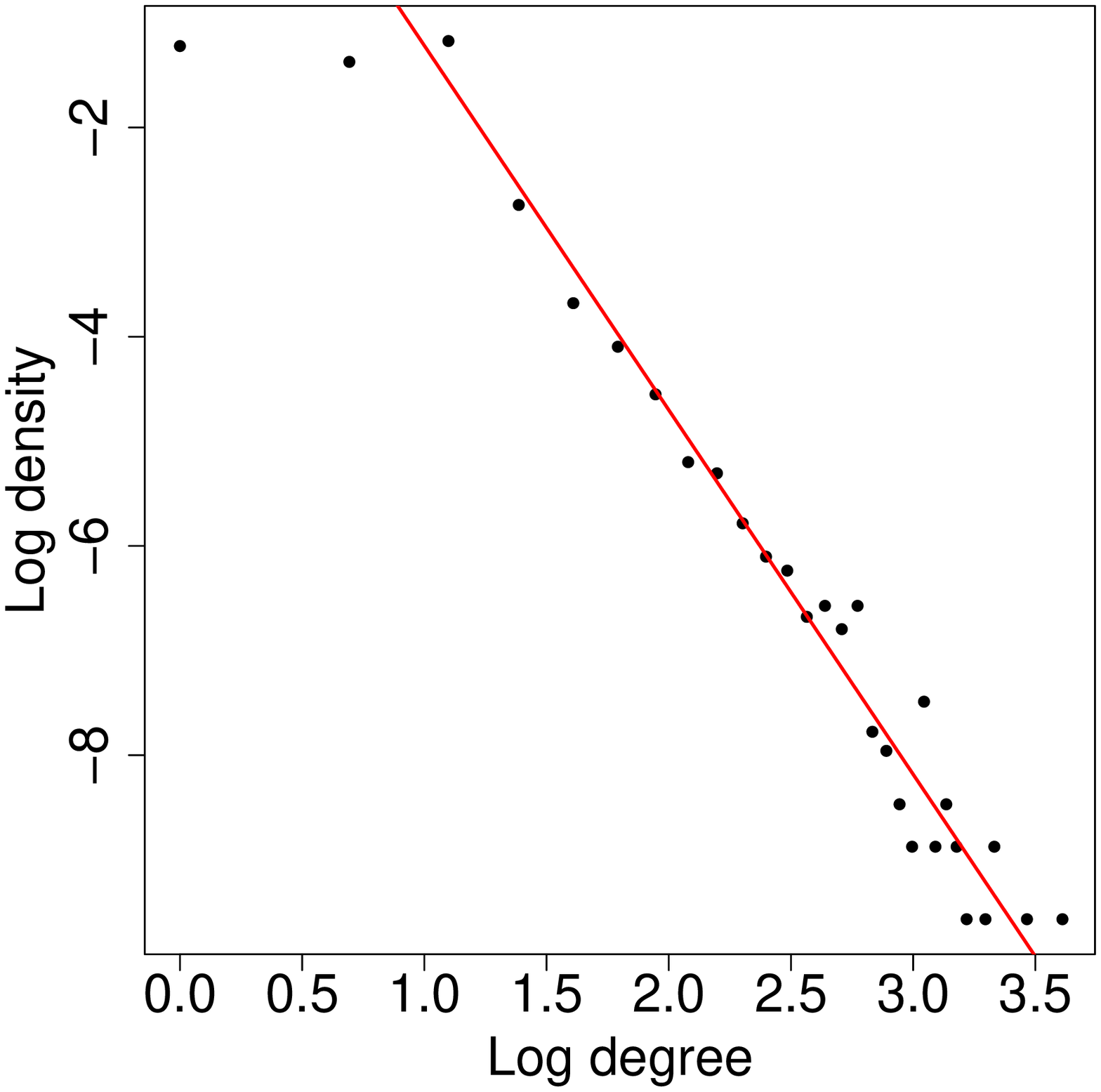}
%\vspace*{-0.2cm}
\caption{$G_{WI}$}
%\label{fig:deg_dist_SERC}
\vspace*{0.2cm}
\end{subfigure}
\begin{subfigure}[b]{0.24\textwidth}
\vspace*{-0.2cm}
\includegraphics[scale=0.25]{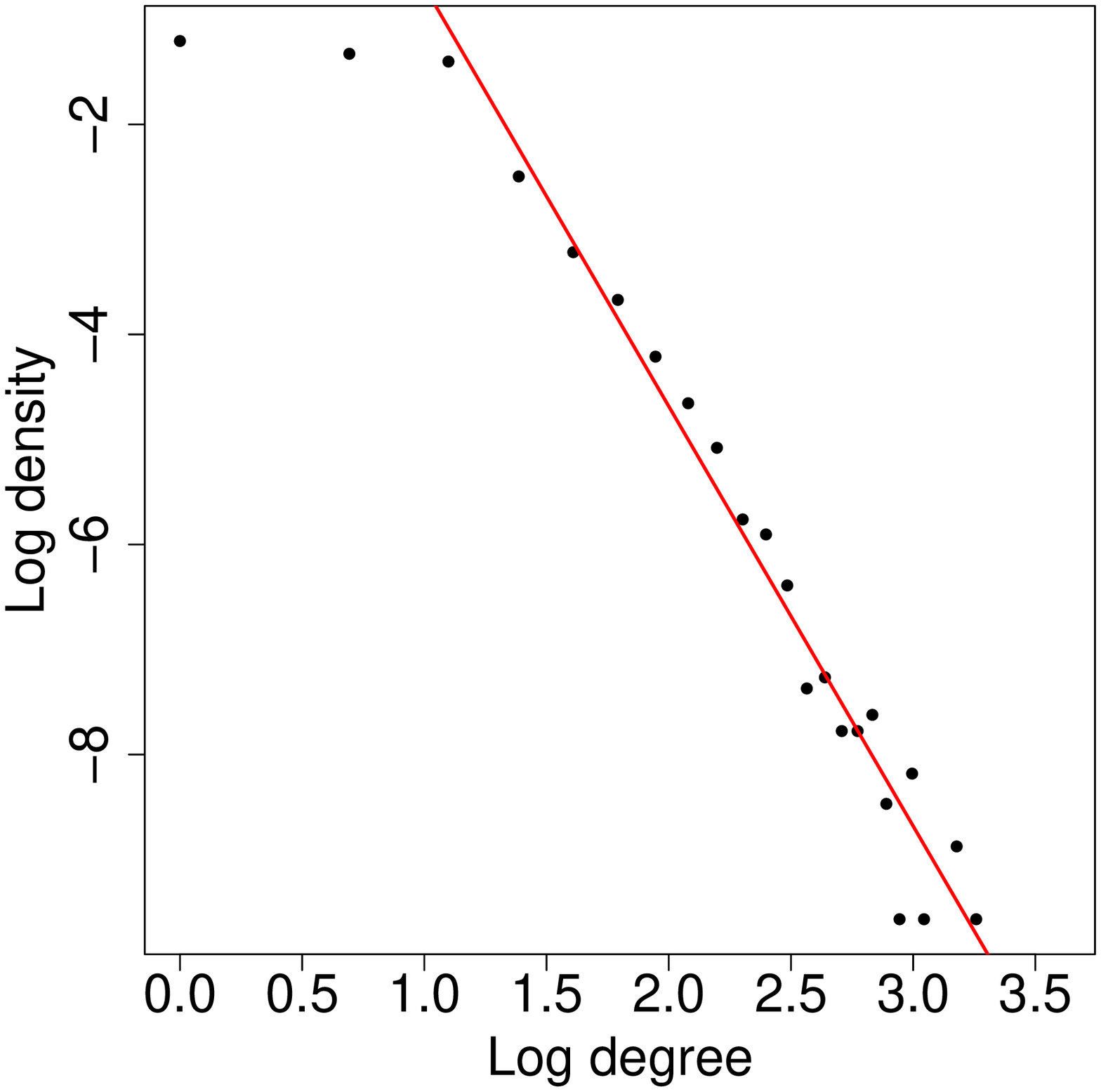}
%\vspace*{-0.2cm}
\caption{$G_{WI}'$}
%\label{fig:deg_dist_gen_EI}
\vspace*{0.2cm}
\end{subfigure}
\vspace*{-0.7cm}
\caption{The degree distribution of the nodes in $G_{WI}$ and $G_{WI}'$ (in
log-log scale). Linear regression lines with slopes $\zeta=-3.48$ and $\zeta=-3.99$ are fitted  to
the distributions of the nodes with degree greater that 2 in $G_{WI}$ and $G_{WI}'$, respectively. The KS statistic between the degree distributions is $0.047$.}
\label{fig:deg_dist_gen_WI}
\vspace*{0.2cm}
\end{figure}
\begin{figure}[t]
\centering
\vspace*{-0.2cm}
\begin{subfigure}[b]{0.23\textwidth}
\vspace*{-0.2cm}
\includegraphics[scale=0.22]{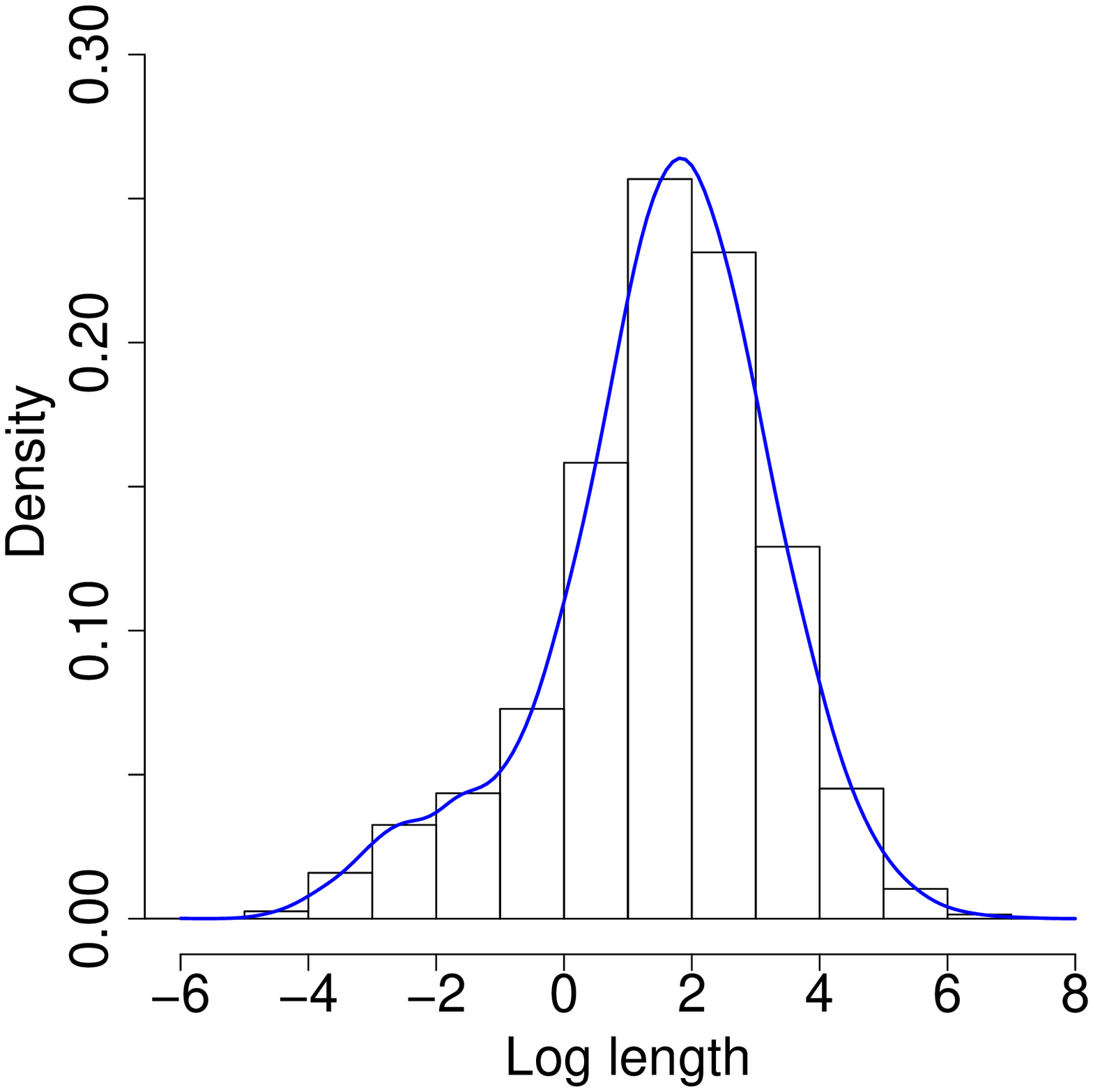}
\caption{$G_{WI}$}
\vspace*{0.2cm}
\end{subfigure}
\begin{subfigure}[b]{0.23\textwidth}
\vspace*{-0.2cm}
\includegraphics[scale=0.22]{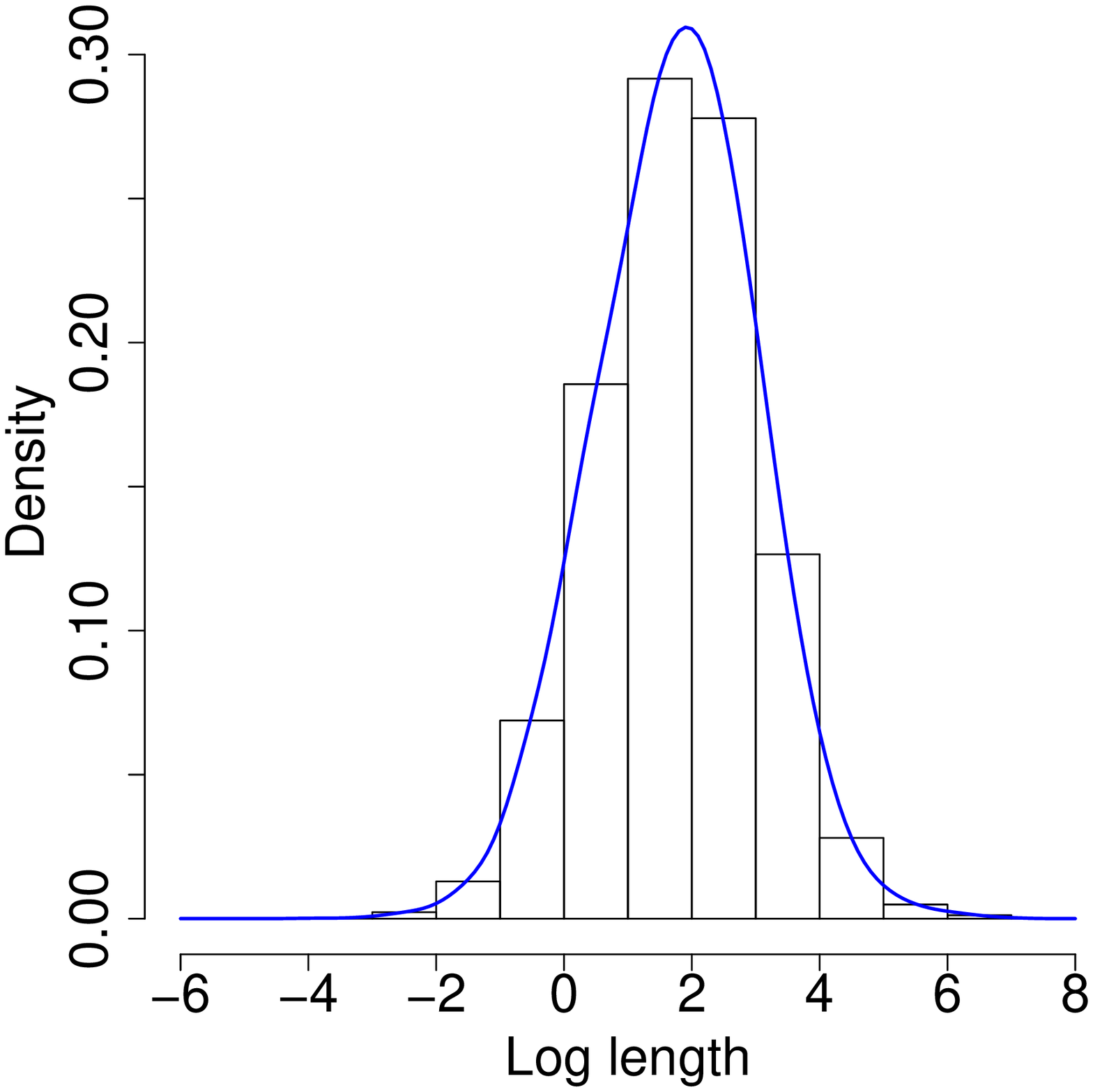}
%\vspace*{-0.2cm}
\caption{$G'_{WI}$}
\vspace*{0.2cm}
\end{subfigure}
\vspace{-0.4cm}
\caption{The length (in \emph{km}) distribution of the point-to-point lines in  $G_{WI}$ and $G_{WI}'$ and nonparametric distribution fit (shown in blue). The KL-divergence between the length distributions in $G_{WI}$ and $G_{WI}'$ is 0.14.}
\label{fig:dist_lines_gen_WI}
\vspace*{0.2cm}
\end{figure}

 Fig.~\ref{fig:deg_dist_gen_WI} shows the degree distribution of the nodes in $G_{WI}'$. As can be seen, the slope of the fitted regression line to the tail of the distribution is $-3.99$ which is similar to that of $G_{WI}$ ($-3.4$). Moreover, the KS statistic between the cumulative degree distributions in $G_{WI}$ and $G_{WI}'$ is $0.047$, indicating the similarity between the degree distributions.
Fig.~\ref{fig:dist_lines_gen_WI} shows the length distribution of the lines in $G_{WI}'$. Since the GNLG Algorithm uses straight lines to connect the nodes, we compare the length distribution of the lines in $G_{WI}'$ with the length distribution of the straight point-to-point lines in $G_{WI}$.  The KL-divergence between the length distributions of the lines in $G_{WI}$ and $G_{WI}'$ is $D_{KL}=0.14$, indicating that  distributions are similar.

 Table~\ref{tb:summary_WI} summarizes the structural properties of the $G_{WI}$ and five instances generated by the GNLG Algorithm. The results indicate that the Algorithm can generate synthetic networks with similar structural properties to the WI grid.

\begin{table}[t]
\centering
\vspace*{-.1cm}
\caption{Comparison between the structural properties of WI ($G_{WI}$) and the Generated WI ($G'_{WI}$). Five instances of $G'_{WI}$ are shown to illustrate that the metric values are similar. All networks have 14,302 nodes and 18,769 edges.}
\vspace*{-0.2cm}
\footnotesize
\begin{tabular}{|l|c|c|c|c|c|}
\hline
Networks& $L$&$C$& $\zeta$  & $D_{KS}$& $D_{KL}$\\
\hline
$G_{WI}$ & 17.33 &0.049 & -3.48 & 0&0\\
\hline
$G'_{WI}$ & 17.40 & 0.052 & -3.99 &0.047& 0.14\\
\hline
$G'_{WI} (2)$ & 18.36 & 0.052 & -3.65 &0.050& 0.15\\
\hline
$G'_{WI} (3)$ & 18.36 & 0.049 & -3.99 &0.047& 0.12\\
\hline
$G'_{WI} (4)$ & 19.06 & 0.052 & -3.61 &0.049& 0.14\\
\hline
$G'_{WI} (5)$ & 17.79 & 0.051 & -3.50&0.049 & 0.14\\
\hline
\end{tabular}
\label{tb:summary_WI}
\end{table}
\normalsize

 \subsection{SERC}
\begin{figure}[t]
\centering
\vspace*{-0.2cm}
\includegraphics[scale=0.4]{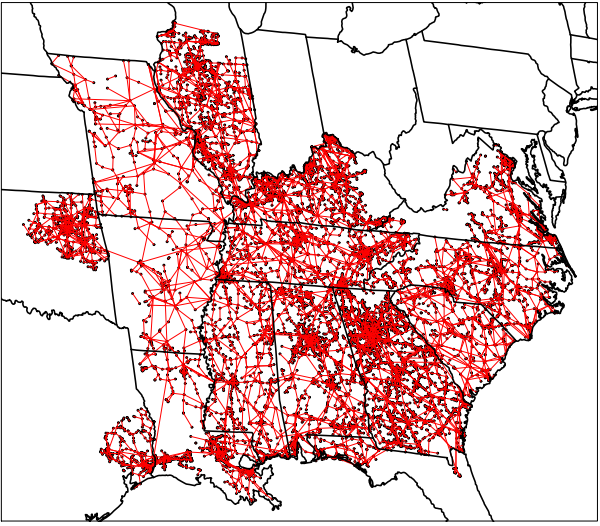}
\vspace*{-0.2cm}
\caption{A part of the Eastern Interconnection (EI) with  12,946 substations (nodes) and 16,658 lines (edges) that operates under the SERC.}
\label{fig:SERC}
\vspace*{0.2cm}
\end{figure}

\begin{figure}[t]
\centering
\vspace*{-0.2cm}
\includegraphics[scale=0.4]{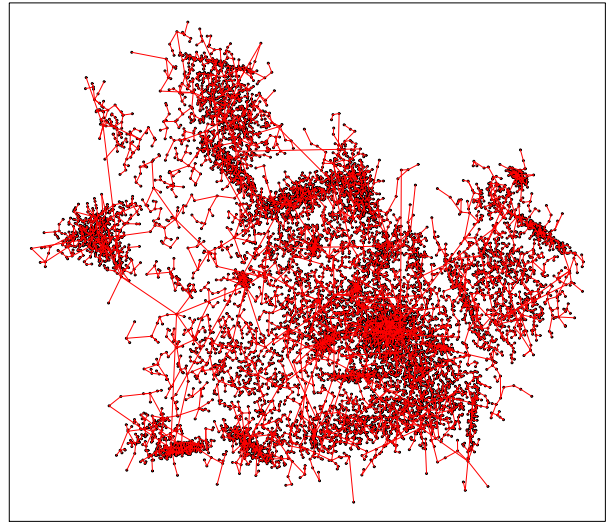}
\vspace*{-0.2cm}
\caption{A network with 12,946 nodes and 16,658 edges generated based on the SERC grid using the GNLG Algorithm with $\kappa=3$, $\alpha = 0.5, \beta = 3.2, \gamma = 2.5,$ and $N=5$.}
\label{fig:Generated_SERC}
\vspace*{0.2cm}
\end{figure}

We apply the GNLG Algorithm to part of the EI that operates under the SERC (see Fig.~\ref{fig:SERC}) that has 13,602 substations (nodes) and 17,767 lines (edges) . Fig.~\ref{fig:Generated_SERC} shows the obtained network using the GNLG Algorithm with $\kappa=3$, $\alpha = 0.5, \beta = 3.2, \gamma = 2.5$,  and $N=5$ that are selected empirically  following several numerical experiments. In the SDNG Procedure, SERC has been clustered into $c=50$ clusters based on the BIC. %(see Fig.~\ref{fig:BIC_SERC}).

\begin{figure}[t]
\centering
\begin{subfigure}[b]{0.24\textwidth}
\vspace*{-0.2cm}
\includegraphics[scale=0.25]{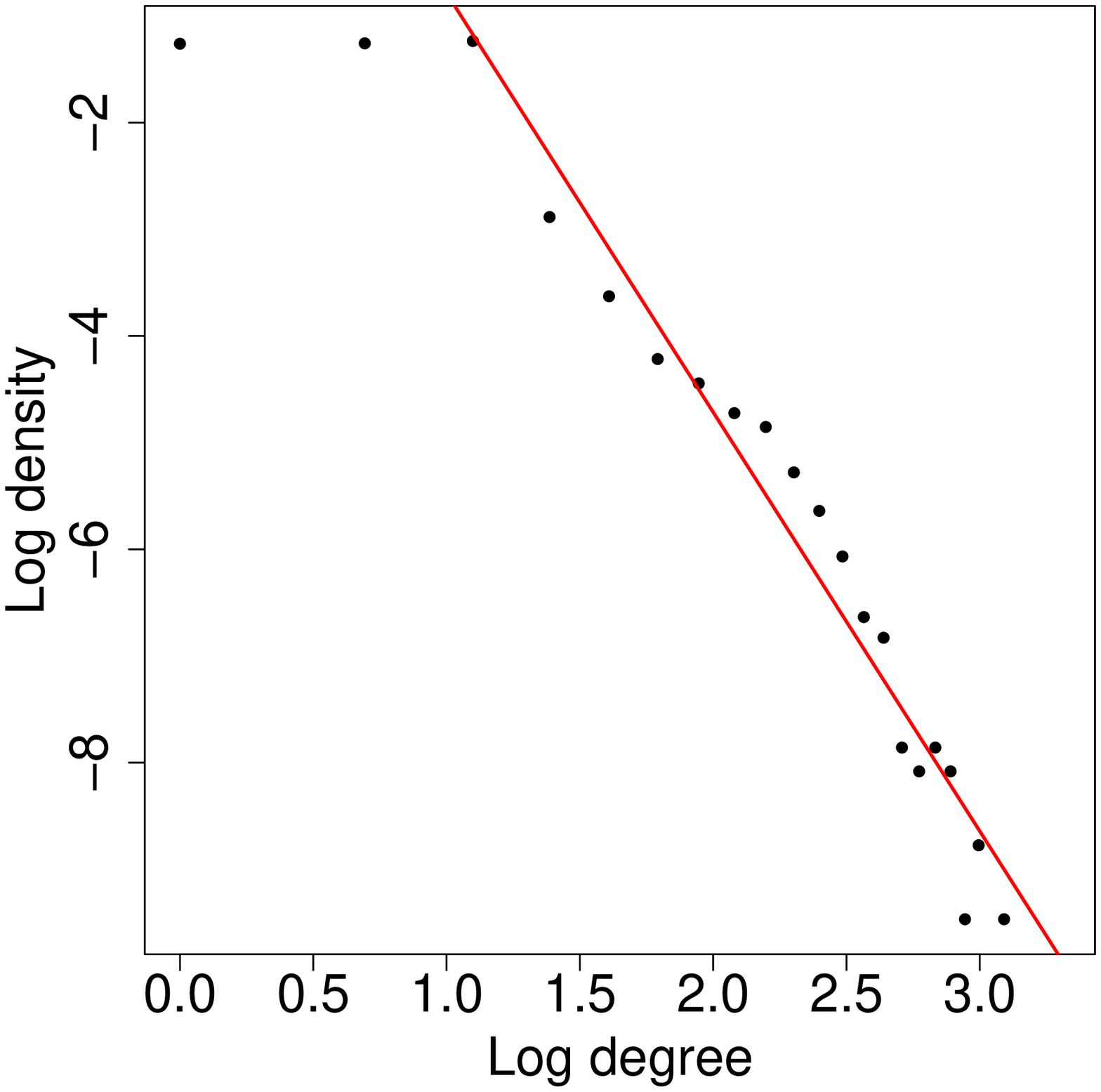}
\caption{$G_{SERC}$}
\vspace*{0.2cm}
\end{subfigure}
\begin{subfigure}[b]{0.24\textwidth}
\vspace*{-0.2cm}
\includegraphics[scale=0.25]{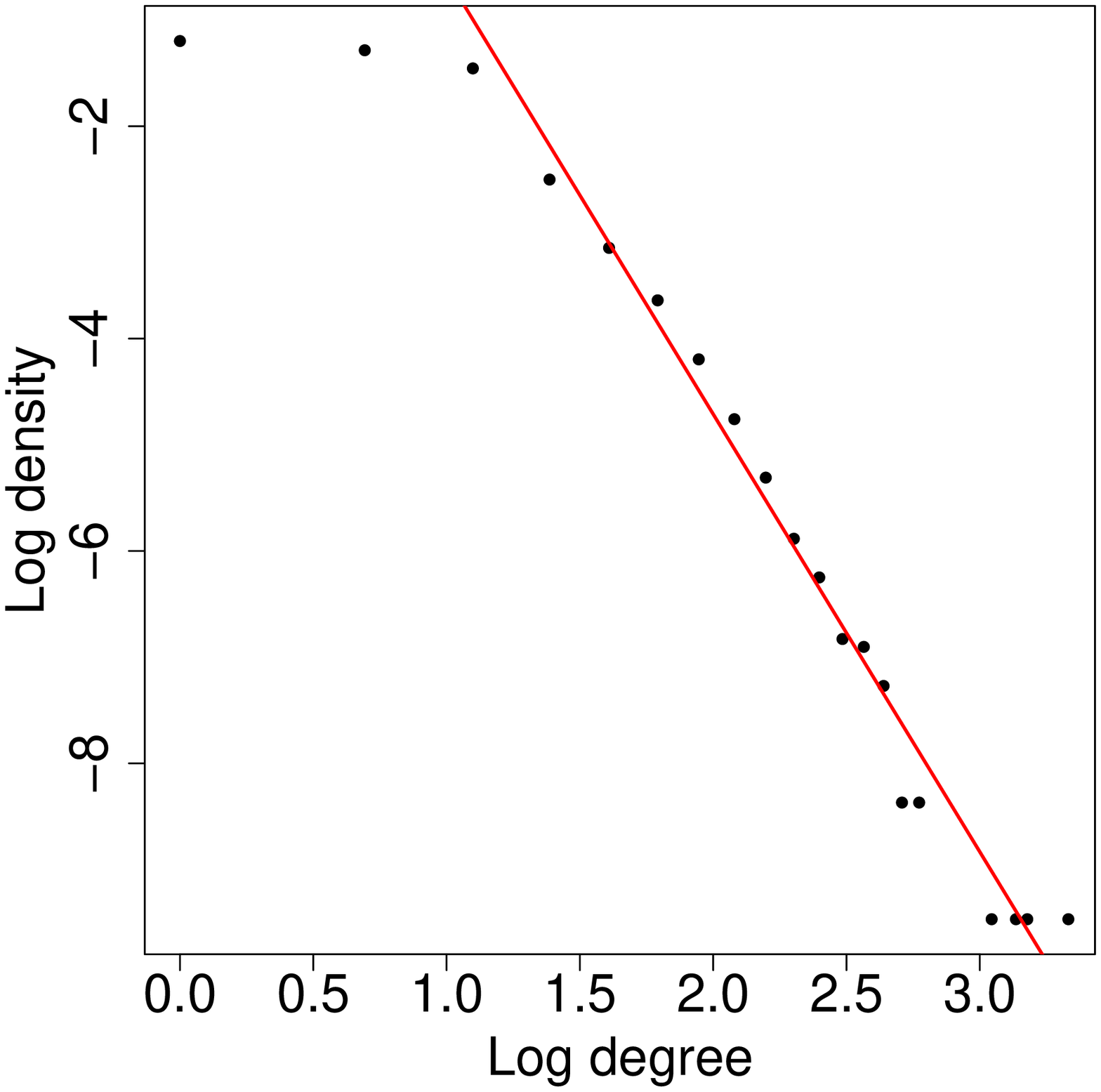}
%\vspace*{-0.2cm}
\caption{$G_{SERC}'$}
\vspace*{0.2cm}
\end{subfigure}
\vspace{-0.6cm}
\caption{The degree distribution of the nodes in $G_{SERC}$ and $G_{SERC}'$ (in
log-log scale). Linear regression lines with slopes $\zeta=-3.93$ and $\zeta=-4.12$ are fitted  to
the distribution of the nodes with degree greater that 2 in $G_{SERC}$ and $G_{SERC}'$, respectively. The KS statistic between the degree distributions is $0.047$.}
\label{fig:deg_dist_SERC}
\end{figure}

\begin{figure}[t]
\centering
\begin{subfigure}[b]{0.23\textwidth}
\vspace*{-0.2cm}
\includegraphics[scale=0.22]{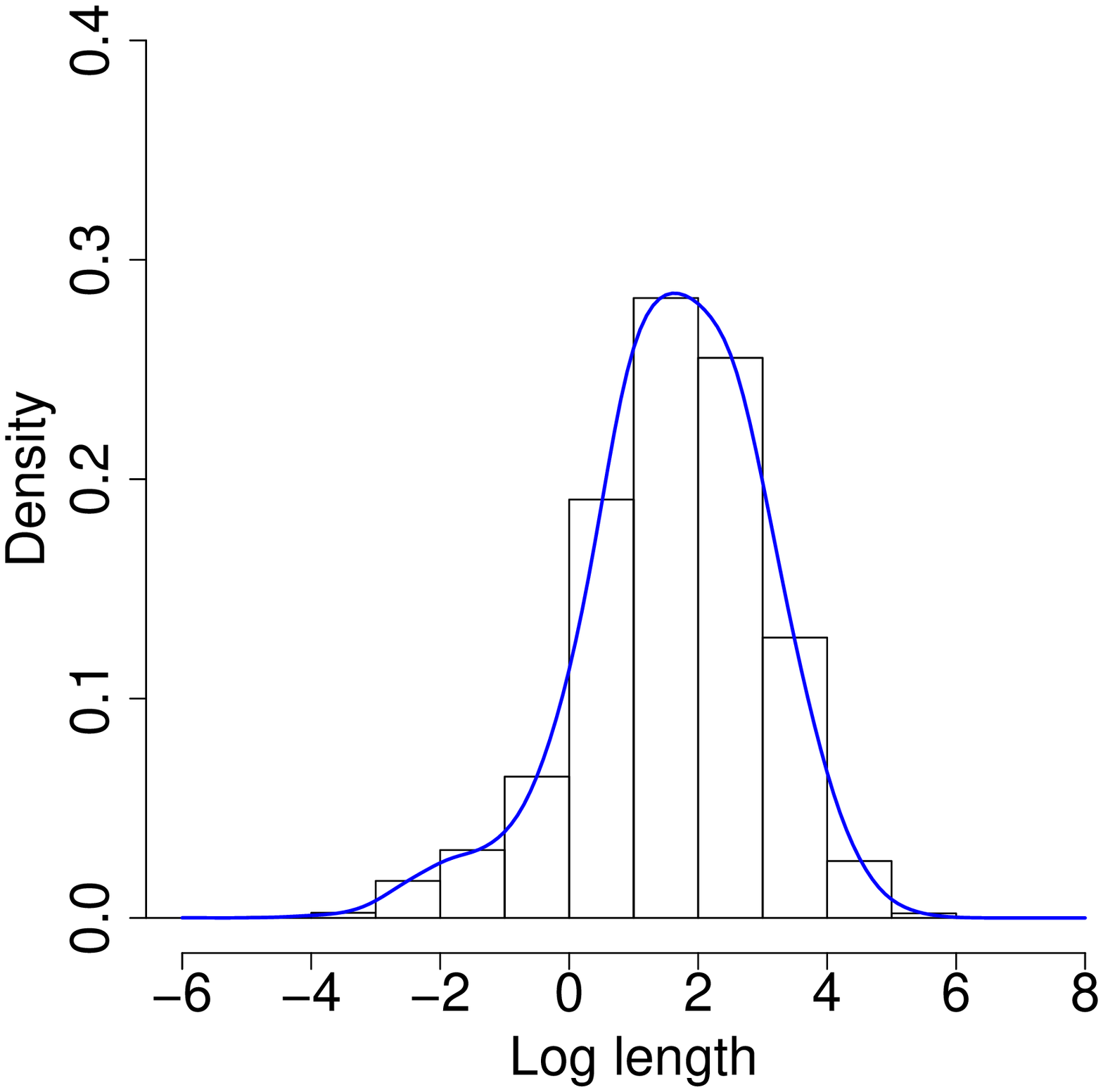}
\caption{$G_{SERC}$}
\vspace*{0.2cm}
\end{subfigure}
\begin{subfigure}[b]{0.23\textwidth}
\vspace*{-0.2cm}
\includegraphics[scale=0.22]{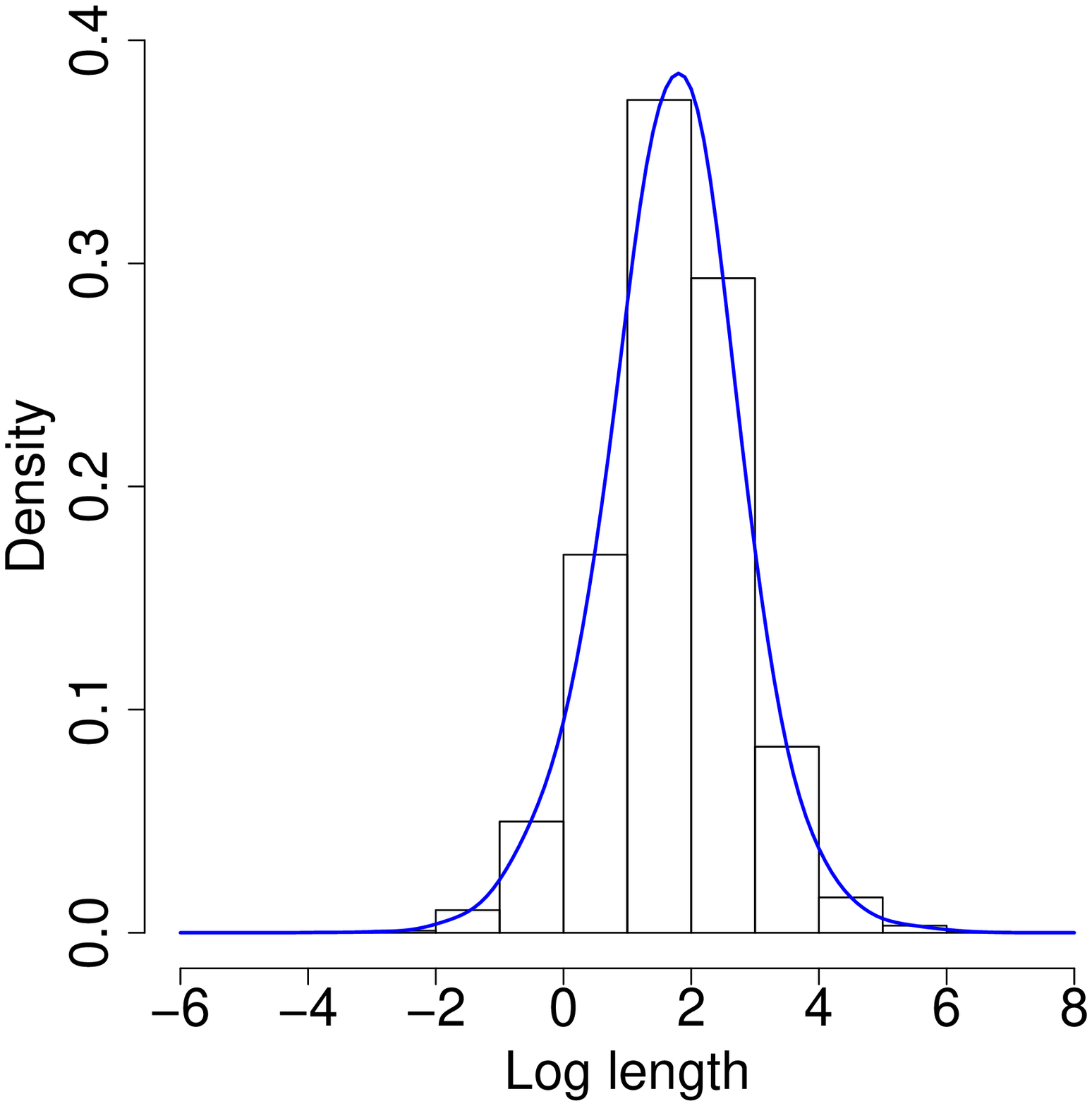}
%\vspace*{-0.2cm}
\caption{$G'_{SERC}$}
\vspace*{0.2cm}
\end{subfigure}
\vspace{-0.2cm}
\caption{The  length (in \emph{km}) distribution of the point-to-point lines in  $G_{SERC}$ and $G_{SERC}'$ and nonparametric distribution fit (shown in blue). The KL-divergence between the length distribution of the lines in $G_{SERC}$ and $G_{SERC}'$ is 0.081.}
\label{fig:dist_lines_SERC}
\end{figure}

The comparison between the degree distribution of the nodes and the length distributions of the lines in $G_{SERC}$ and $G'_{SERC}$ are shown in Figs.~\ref{fig:deg_dist_SERC} and \ref{fig:dist_lines_SERC}. Table~\ref{tb:summary_SERC}, summarizes the structural properties of $G_{SERC}$ and five instances generated by the GNLG Algorithm. As with the WI, it can be seen that the Algorithm can generate synthetic networks with similar structural properties to the SERC grid.

\subsection{FRCC}\label{subsec:FRCC}
\begin{figure}[t]
\centering
\begin{subfigure}[b]{0.24\textwidth}
\vspace*{-0.2cm}
\includegraphics[scale=0.25]{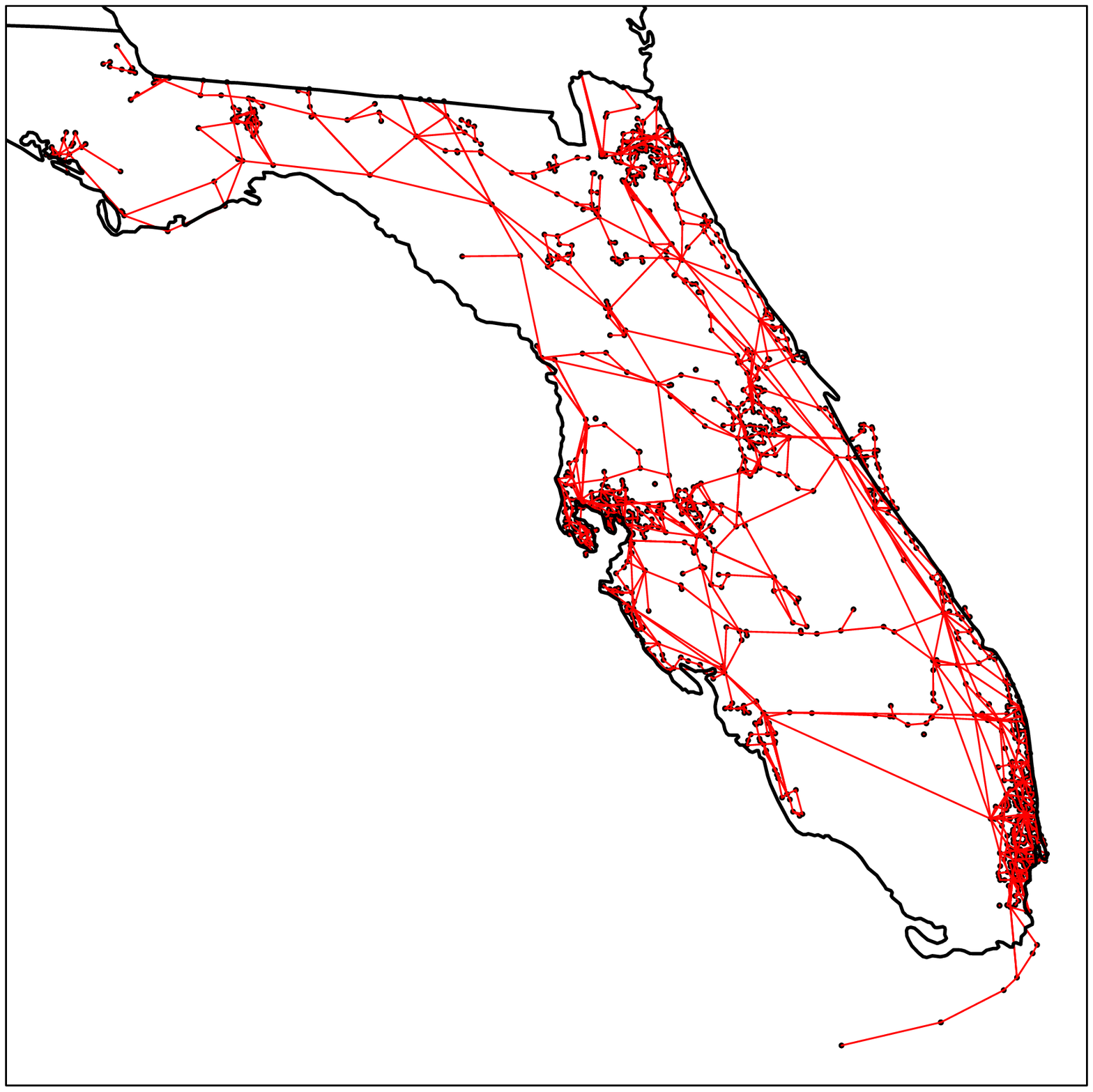}
%\vspace*{-0.2cm}
\caption{$G_{FRCC}$}
%\label{fig:deg_dist_SERC}
\vspace*{0.2cm}
\end{subfigure}
\begin{subfigure}[b]{0.24\textwidth}
\vspace*{-0.2cm}
\includegraphics[scale=0.25]{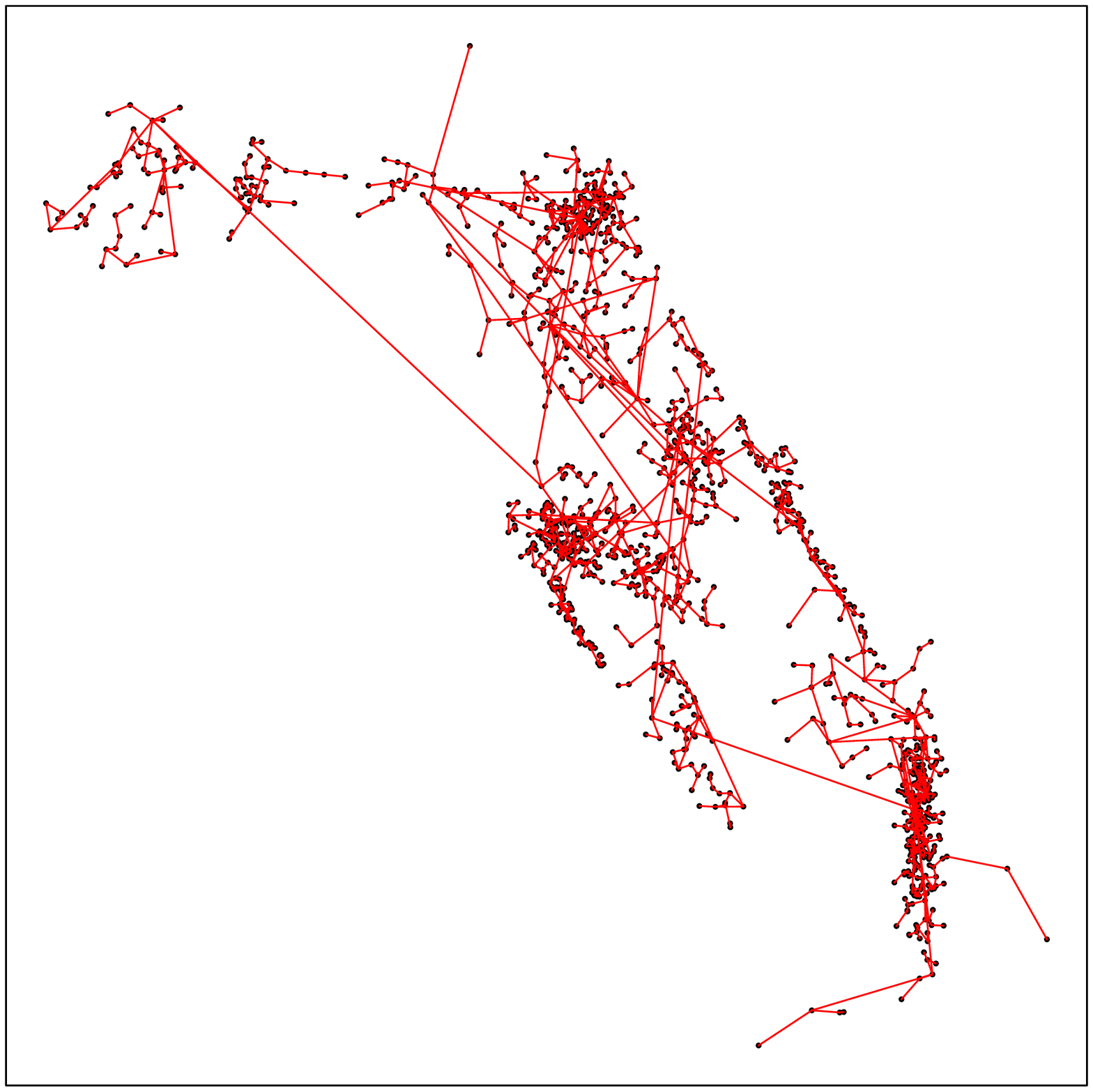}
%\vspace*{-0.2cm}
\caption{$G_{FRCC}'$}
%\label{fig:deg_dist_gen_EI}
\vspace*{0.2cm}
\end{subfigure}
\vspace{-0.4cm}
\caption{(a) Part of the  Eastern Interconnection (EI)  with  1,312 substations (nodes) and 1,780 lines (edges) that operates under the FRCC. (b) A network with the same number of nodes and edges that is generated using the GNLG Algorithm with $\kappa=1.8, \alpha = 0.5, \beta = 2.5, \gamma = 2.8,$ and $N=5$.}
\label{fig:FRCC}
\end{figure}

\begin{figure}[t]
\centering
\begin{subfigure}[b]{0.24\textwidth}
\vspace*{-0.2cm}
\includegraphics[scale=0.25]{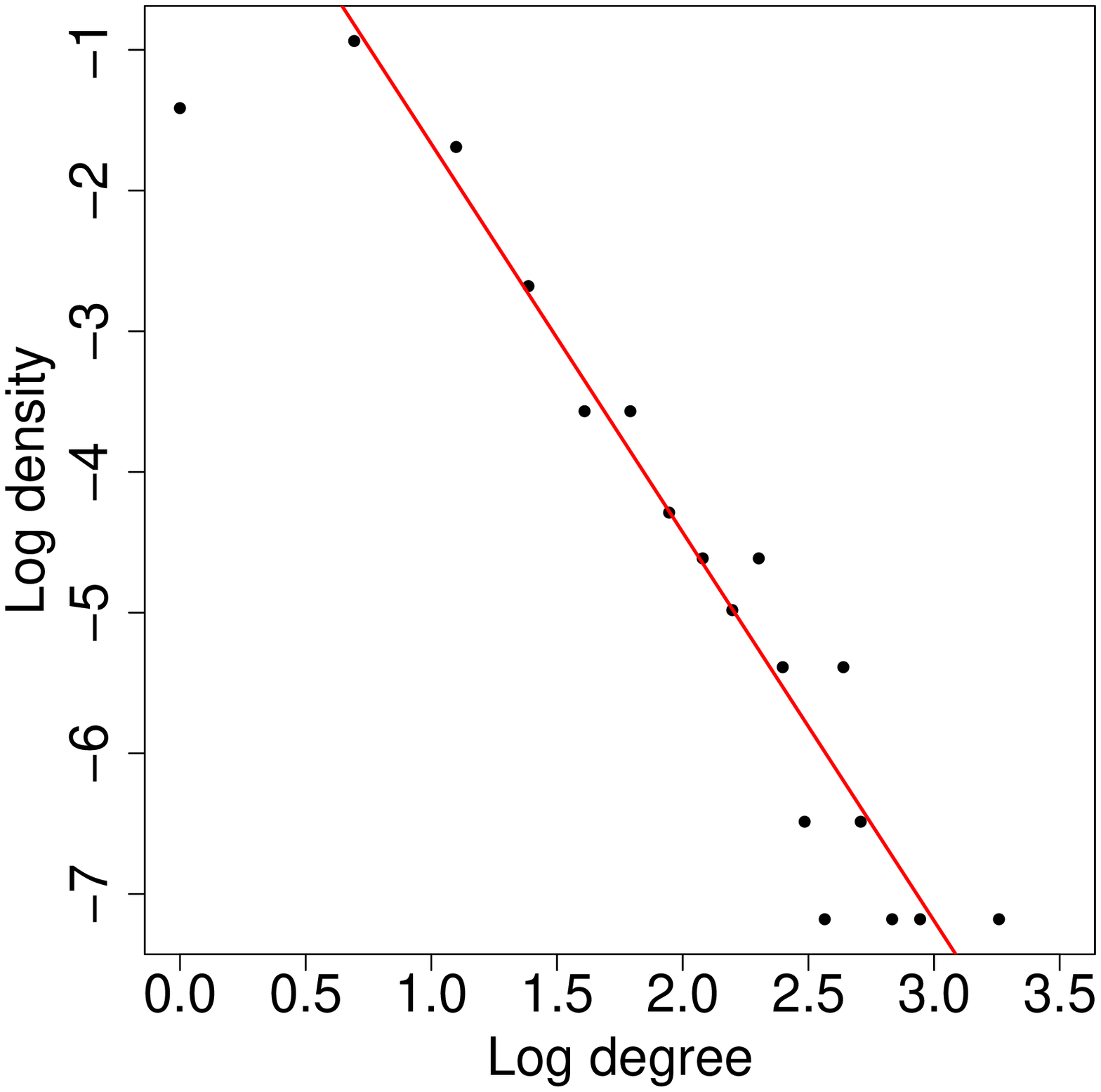}
%\vspace*{-0.2cm}
\caption{$G_{FRCC}$}
%\label{fig:deg_dist_SERC}
\vspace*{0.2cm}
\end{subfigure}
\begin{subfigure}[b]{0.24\textwidth}
\vspace*{-0.2cm}
\includegraphics[scale=0.25]{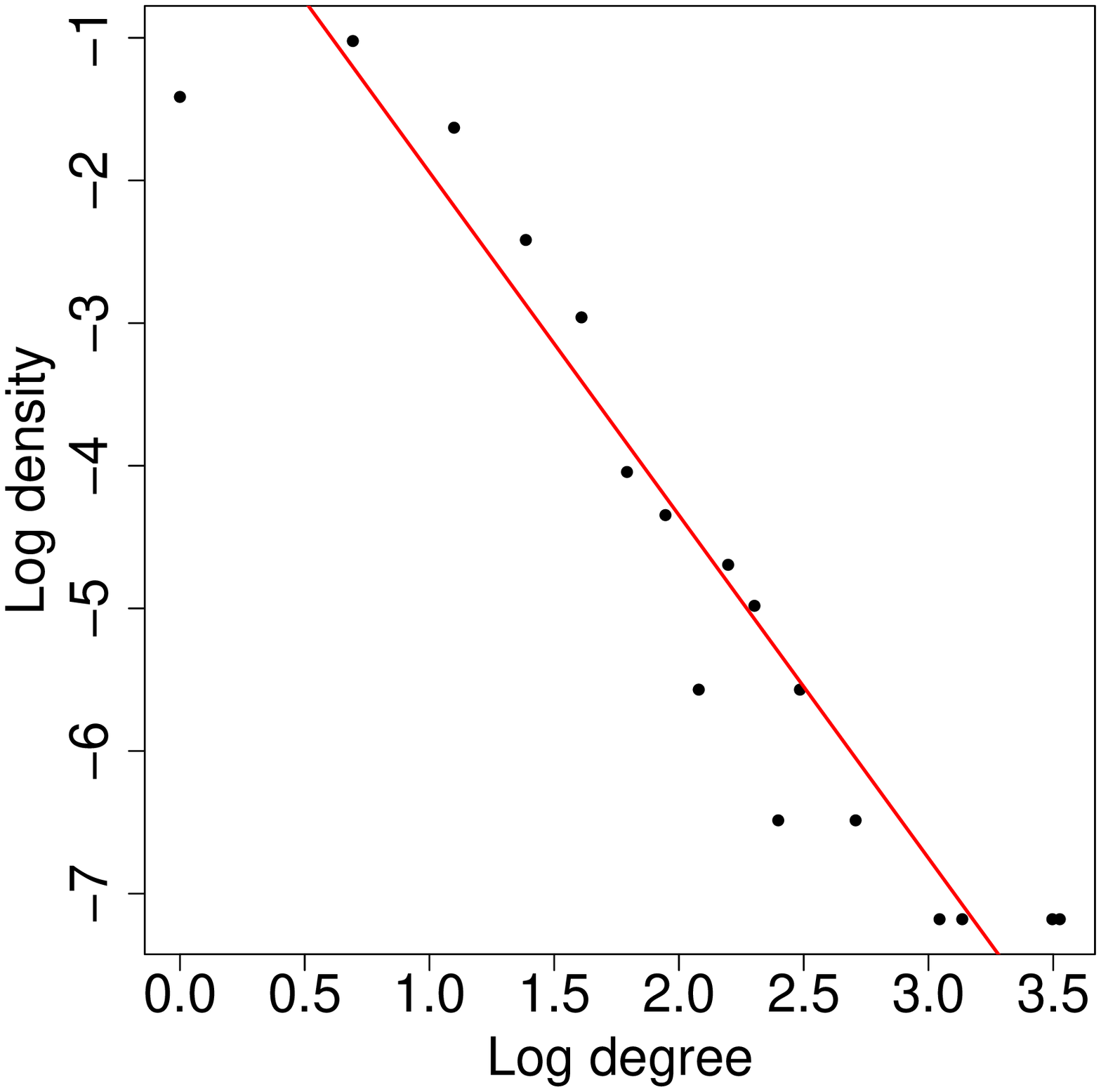}
%\vspace*{-0.2cm}
\caption{$G_{FRCC}'$}
%\label{fig:deg_dist_gen_EI}
\vspace*{0.2cm}
\end{subfigure}
\vspace{-0.6cm}
\caption{The degree distribution of the nodes in $G_{FRCC}$ and $G_{FRCC}'$ (in
log-log scale). Linear regression lines with slopes $\zeta=-2.76$ and $\zeta=-2.40$ are fitted  to
the distribution of the nodes with degree greater that 1 in $G_{FRCC}$ and $G_{FRCC}'$, respectively. The KS statistic between the degree distributions is $0.032$.}
\label{fig:deg_dist_FRCC}
\end{figure}

Finally, we apply the GNLG Algorithm to a smaller part of the EI with  1,312 substations (nodes) and 1,780 lines (edges) that operates under the FRCC (see Fig.~\ref{fig:FRCC}).  As can be seen in Fig.~\ref{fig:deg_dist_FRCC}, the degree distribution of the nodes in $G_{FRCC}$ is different from the degree distribution of the nodes in $G_{WI}$ and $G_{SERC}$. In $G_{FRCC}$, only the density of the nodes with degree 1 is not on the fitted regression line. This suggests that in the Reinforcement Procedure, the step for small networks should be used and nodes should be sampled with probability $\propto d_i'^{-\eta}\rho_i'^{-\alpha}$. Here, we use $\eta=2$.

\begin{figure}[t]
\centering
\begin{subfigure}[b]{0.23\textwidth}
\vspace*{-0.2cm}
\includegraphics[scale=0.22]{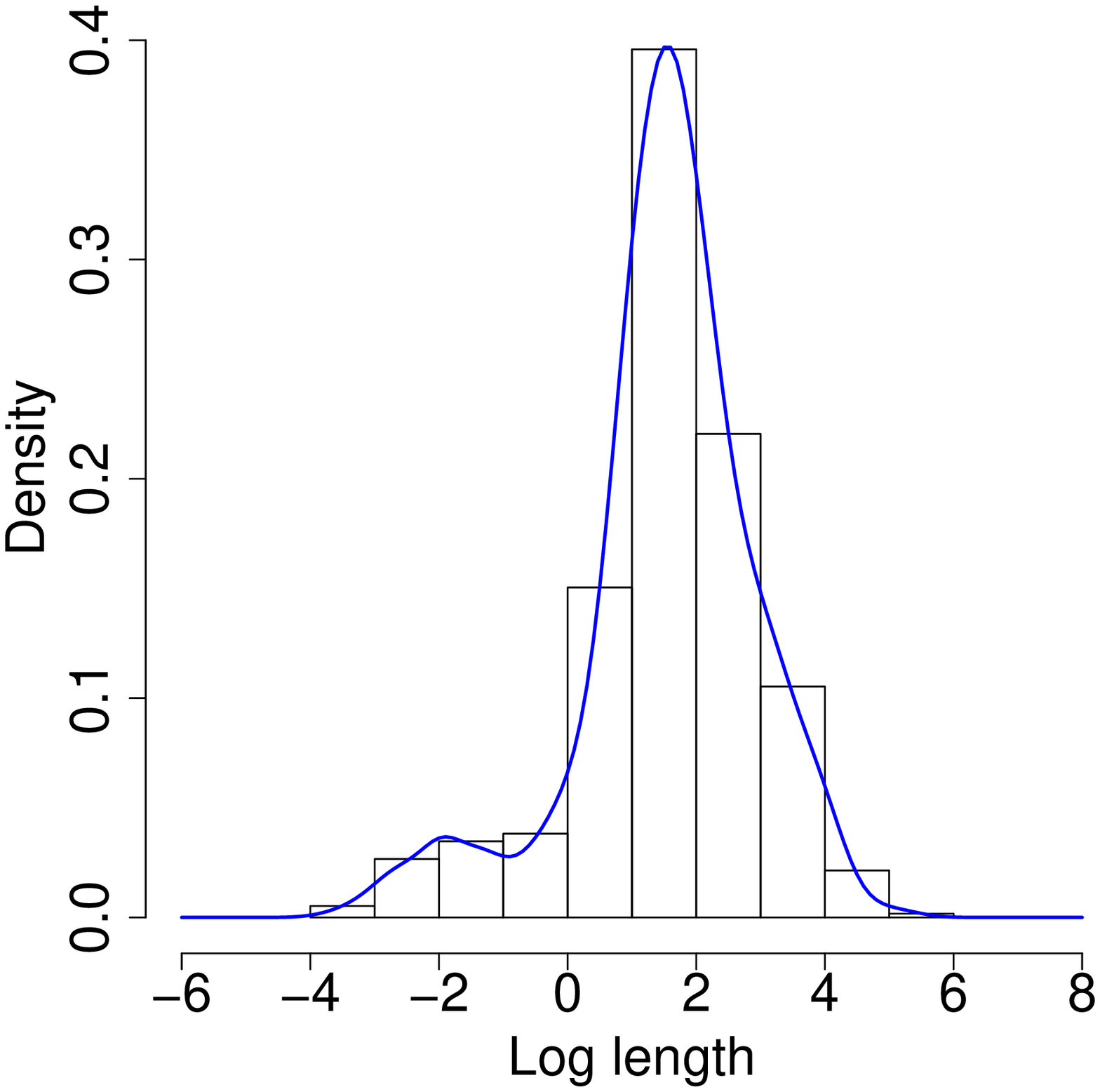}
%\vspace*{-0.2cm}
\caption{$G_{FRCC}$}
%\label{fig:dist_lines_EI}
\vspace*{0.2cm}
\end{subfigure}
\begin{subfigure}[b]{0.23\textwidth}
\vspace*{-0.2cm}
\includegraphics[scale=0.22]{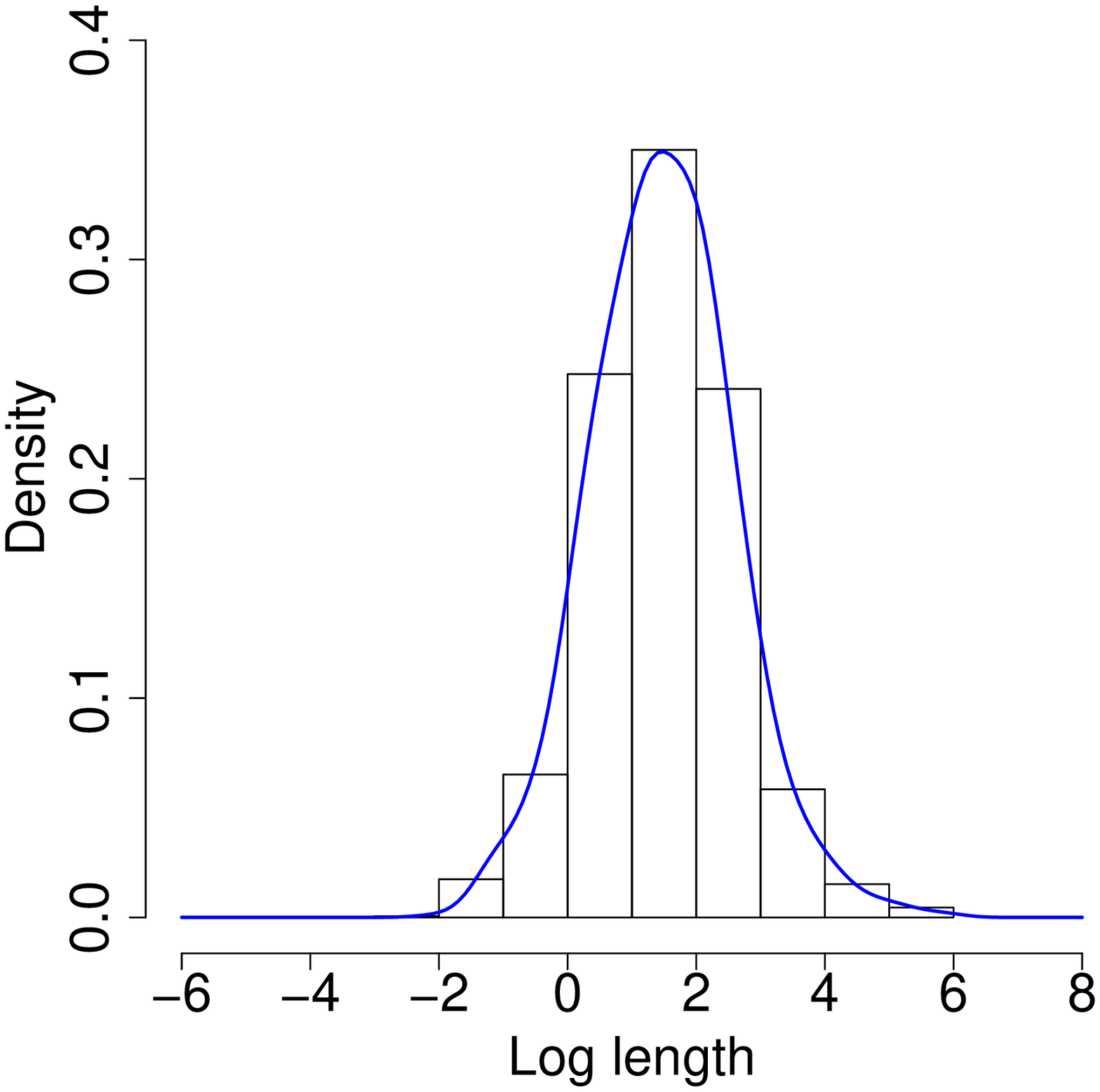}
%\vspace*{-0.2cm}
\caption{$G'_{FRCC}$}
%\label{fig:dist_lines_gen_EI}
\vspace*{0.2cm}
\end{subfigure}
\vspace{-0.2cm}
\caption{The  length (in \emph{km}) distribution of the point-to-point lines in  $G_{FRCC}$ and $G_{FRCC}'$ and nonparametric distribution fit (shown in blue). The KL-divergence between the length distributions in $G_{FRCC}$ and $G_{FRCC}'$ is 0.12.}
\label{fig:dist_lines_FRCC}
\end{figure}

Fig.~\ref{fig:FRCC} shows the obtained network using the GNLG Algorithm with $\kappa=1.8, \alpha = 0.5, \beta = 2.5, \gamma = 2.8,$ and $N=5$ that were selected empirically. Nodes in the FRCC has been clustered into $c=15$ clusters. The comparison between the degree distributions of the nodes and length distributions of the lines between $G_{FRCC}$ and in $G'_{FRCC}$ are shown in Figs.~\ref{fig:deg_dist_FRCC} and \ref{fig:dist_lines_FRCC}. Table~\ref{tb:summary_FRCC}, summarizes the structural properties of the FRCC and five instances generated by the GNLG Algorithm. The results suggest that the GNLG algorithm can generate smaller networks as well.

\begin{table}[t]
\centering
\vspace*{-.1cm}
\caption{Comparison between the structural properties of the SERC ($G_{SERC}$) and the Generated SERC ($G'_{SERC}$). Five instances are shown to illustrate that the metric values are similar. All networks have 12,946 nodes and 16,658 edges.}
\vspace*{-0.2cm}
\footnotesize
\begin{tabular}{|l|c|c|c|c|c|}
\hline
Networks& $L$&$C$& $\zeta$  & $D_{KS}$& $D_{KL}$\\
\hline
$G_{SERC}$ & 19.71 &0.049 & -3.93 & 0&0\\
\hline
$G'_{SERC}$ & 20.26 & 0.048 & -4.12 &0.047& 0.081\\
\hline
$G'_{SERC} (2)$ & 19.43 & 0.045 & -4.25 &0.044& 0.077\\
\hline
$G'_{SERC} (3)$ & 17.56 & 0.048 & -4.72 &0.044& 0.084\\
\hline
$G'_{SERC} (4)$ & 17.95 & 0.047 & -4.46 &0.048& 0.083\\
\hline
$G'_{SERC} (5)$ & 19.87 & 0.049 & -4.5 & 0.046&0.080\\
\hline
\end{tabular}
\label{tb:summary_SERC}
\end{table}
\normalsize

\begin{table}[t]
\centering
\vspace*{-.1cm}
\caption{Comparison between the structural properties of the FRCC ($G_{FRCC}$) and the Generated FRCC ($G'_{FRCC}$). Five instances are shown to illustrate that the metric values are similar. All networks have 1,312 nodes and 1,780 edges.}
\vspace*{-0.2cm}
\footnotesize
\begin{tabular}{|l|c|c|c|c|c|}
\hline
Networks& $L$&$C$& $\zeta$ & $D_{KS}$& $D_{KL}$\\
\hline
$G_{FRCC}$ & 11.68 &0.075 & -2.76 & 0&0\\
\hline
$G'_{FRCC}$ & 10.81 & 0.045 & -2.40&0.032 & 0.12\\
\hline
$G'_{FRCC} (2)$ & 11.86 & 0.057 & -2.70 &0.025& 0.12\\
\hline
$G'_{FRCC} (3)$ & 11.13 & 0.053 & -2.78 &0.022& 0.10\\
\hline
$G'_{FRCC} (4)$ & 11.27 & 0.051 & -2.86 &0.025& 0.13\\
\hline
$G'_{FRCC} (5)$ & 11.66 & 0.057 & -2.36 &0.015& 0.12\\
\hline
\end{tabular}
\label{tb:summary_FRCC}
\end{table}
\normalsize
\section{Conclusions}\label{sec:conclusion}

In this paper, we developed the GNLG Algorithm for generating synthetic power grid networks with similar structural properties to a given network. We applied the algorithm to the WI and two parts of the EI (SERC and FRCC) and showed that it can generate networks with similar structural properties to these networks. In a broader perspective, the algorithm can be used for anonymizing network data that cannot be published otherwise, thereby enabling research in power grid vulnerability and resilience.

This is only a first step towards generation of synthetic power grid networks and there are clearly several future research directions. Specifically, for a given network, step 1 of the GNLG Algorithm and tuning the parameters need to be done only once. Then, the algorithm can be used to generate several networks similar to a given network.
Hence, we plan to provide a web application that would allow obtaining synthetic networks similar to a given reliability regions in the Northern American power grid with specific set of parameters (e.g., currently it takes less than 3.5 minutes for our server to generate a synthetic network similar to the WI). Moreover, we plan to improve the algorithm and to focus on locations of power generators and demand nodes as well as on generation and demand values. Generation of topologies where the line voltages are taken into account is also an interesting open problem.
Finally, we believe that the approach can be extended for generating various types of spatially distributed networks.

\section*{Acknowledgement}
This work was supported in part by DTRA grant HDTRA1-13-1-0021, CIAN NSF ERC under grant EEC-0812072, the People Programme (Marie Curie Actions) of the European Unions Seventh Framework Programme (FP7/2007-2013) under REA grant agreement no. [PIIF-GA-2013-629740].11.

\bibliographystyle{IEEEtran}
\bibliography{bib}

\end{document}